\documentclass{jfm}

\usepackage{graphicx}
\usepackage{natbib}
\usepackage[utf8]{inputenc}
\usepackage[T1]{fontenc}
\usepackage{graphicx}
\usepackage{amsmath}
\usepackage{amsfonts}
\usepackage{amssymb}
\usepackage{caption}
\usepackage{subcaption}
\usepackage{epstopdf}

\DeclareGraphicsExtensions{.pdf,.png,.eps}

\ifCUPmtlplainloaded \else
  \checkfont{eurm10}
  \iffontfound
    \IfFileExists{upmath.sty}
      {\typeout{^^JFound AMS Euler Roman fonts on the system,
                   using the 'upmath' package.^^J}%
       \usepackage{upmath}}
      {\typeout{^^JFound AMS Euler Roman fonts on the system, but you
                   dont seem to have the}%
       \typeout{'upmath' package installed. JFM.cls can take advantage
                 of these fonts,^^Jif you use 'upmath' package.^^J}%
       \providecommand\upi{\pi}%
      }
  \else
    \providecommand\upi{\pi}%
  \fi
\fi
\newcommand{\pderiv}[2]{\frac{\partial #1}{\partial #2}}

\ifCUPmtlplainloaded \else
  \checkfont{msam10}
  \iffontfound
    \IfFileExists{amssymb.sty}
      {\typeout{^^JFound AMS Symbol fonts on the system, using the
                'amssymb' package.^^J}%
       \usepackage{amssymb}%

      }{}
  \fi
\fi

\ifCUPmtlplainloaded \else
  \IfFileExists{amsbsy.sty}
    {\typeout{^^JFound the 'amsbsy' package on the system, using it.^^J}%
     \usepackage{amsbsy}}
    {\providecommand\boldsymbol[1]{\mbox{\boldmath $##1$}}}
\fi

 \renewcommand{\i}{\textup{i}}
\newcommand{\e}{\textup{e}}

\title[Aerodynamic noise from edges with porous extensions]{Aerodynamic noise from rigid trailing edges with finite porous extensions}

\author[A. Kisil, \& L. J. Ayton]
{A. \ns K\ls I\ls S\ls I\ls L, \thanks{Email address for correspondence: A.Kisil@damtp.cam.ac.uk} and L. \ns J.\ns A\ls Y\ls T\ls O\ls N \thanks{Email address for correspondence: L.J.Ayton@damtp.cam.ac.uk}}

\affiliation{Department of Applied Mathematics and Theoretical Physics, University of Cambridge,
Wilberforce Road, CB3 0WA, UK}

\begin{document}

\maketitle

\begin{abstract}
This paper investigates the effects of finite flat porous extensions to semi-infinite impermeable flat plates in an attempt to control trailing-edge noise through bio-inspired adaptations. Specifically the problem of sound generated by a  gust convecting in uniform mean steady flow scattering off the trailing edge and permeable-impermeable junction is considered. This setup supposes that any realistic trailing-edge adaptation to a blade would be sufficiently small so that the turbulent boundary layer encapsulates both the porous edge and the permeable-impermeable junction, and therefore the interaction of acoustics generated at these two discontinuous boundaries is important. The acoustic problem is tackled analytically through use of the Wiener-Hopf method. A two-dimensional matrix Wiener-Hopf problem arises due to the two interaction points (the trailing edge and the permeable-impermeable junction). 
This paper discusses a new iterative method for solving this matrix Wiener-Hopf equation which extends to further two-dimensional problems in particular those involving analytic terms that exponentially grow in the upper or lower half planes. This method is an extension of the commonly used ``pole removal'' technique and avoids the needs for full matrix factorisation. 
Convergence of this iterative method to an exact solution is shown to be particularly fast when terms neglected in the second step are formally smaller than all other terms retained.
The new method is validated by comparing the iterative solutions for acoustic scattering by a finite impermeable plate against a known solution (obtained in terms of Mathieu functions).
The final acoustic solution highlights the effects of the permeable-impermeable junction on the generated noise, in particular how this junction affects the far-field noise generated by high-frequency gusts by creating an interference to typical trailing-edge scattering. This effect results in partially porous plates predicting a lower noise reduction than fully porous plates when compared to fully impermeable plates.
\end{abstract}

\section{Introduction}
The noise generated by aerodynamic bodies in fluid flow has motivated significant research for decades, ranging from simple analytical models for flat plates in uniform flow with unsteady perturbations \citep{Amiet} to high-fidelity numerical models that predict the noise generated by turbulent interactions with realistic thick aerofoils \citep{JamesJASA,Allampalli,Lock,Kim}.
A particularly important and unavoidable source of aerofoil noise is so-called trailing-edge noise which is generated by turbulence in the boundary layer scattering off the sharp trailing edge of an aerofoil. 
The analytic work of \citet{HoweBook} presents a simplified picture of a line vortex interacting with the trailing edge of a semi-infinite flat plate and predicts the effects of a rigid trailing-edge condition on the acoustic scattering. With this boundary condition being key to the total level of far-field noise generated by turbulence-edge interaction, it is clear why the next steps into the investigation of reducing trailing-edge noise were to consider adapted trailing edge designs, for example a serrated edge \citep{Lyu, OerlemansAIAA}, a rough surface canopy \citep{canopy}, or a porous and/or flexible edge \citep{JP, DNSPE,GeyerPorous}.
These adapted designs are inspired by the silent flight of owls, first discussed by \citep{Graham}, and now with increasing pressure on the aviation industry to reduce aircraft noise \citep{ACARE} are of great interest worldwide.

In order to understand and quickly predict any noise-reduction capabilities of adapted trailing-edge designs we turn to analytic solutions to elucidate the physics of the fluid-structure interactions. Recently \citep{JP} used the Wiener-Hopf technique to obtain an analytic solution capable of predicting the far-field trailing-edge noise from a semi infinite poroelastic plate in uniform flow of characteristic Mach number $M<1$. It is shown that, whilst a rigid impermeable edge has a far-field acoustic power scaling of $M^{5}$ \citep{FWHall}, scalings of $M^{6}$ and $M^{7}$ can be achieved with suitable tuning of the porosity and elasticity of the edge. A further key insight is that porosity is seen to dominate low-frequency noise reduction, whilst elasticity affects high-frequency noise. 
The work of \citet{JP} has recently been extended to consider the effects of a finite leading edge \citep{CavaWolf} although in this case the full plate is poroelastic. Practical application of trailing edge adaptations would be restricted to short extensions to minimise adverse aerodynamic effects, therefore it is sensible to assume the turbulent boundary layer would encounter both rigid and porous sections of a plate. For this reason in this paper we investigate the noise generated by a gust convecting over a semi-infinite rigid plate with finite porous extension. To do so we use the Wiener-Hopf technique. 

The Wiener-Hopf technique \citep{Noble} affords itself to the scattering of sound by edges since these are problems with mixed boundary conditions: a velocity condition on the plate $y=0, x<0$, and a pressure continuity condition behind the plate $y=0, x>0$. With two mixed boundary conditions like these, one constructs a one-dimensional (or scalar) Wiener-Hopf problem requiring the multiplicative factorisation of a scalar kernel function, $K(\alpha)$, into two parts that are analytic in the upper and lower halves of the complex $\alpha$-plane, i.e. $K(\alpha)=K_{+}(\alpha)K_{-}(\alpha)$, where $\pm$ denotes analyticity in the upper/lower half plane respectively. Analytic factorisations are tractable in some cases, such as the Sommerfeld diffraction problem \citep{Sommerfeld}, however in more complicated situations such as the poroelastic plate of \citep{JP} closed-form factorisation can only be found in certain asymptotic limits, and otherwise have to be computed numerically.

In more complicated scattering problems with more than two mixed
boundary conditions, such as a finite elastic strip \citep{ScottElastic}, or a
perforated plate \citep{Abrahamsperf}, one is faced with a matrix kernel to
factorise. The
general question of constructive matrix Wiener--Hopf factorisation is 
open: \citep{history, Constructive_review}. In this paper we are
concerned with a class of  Wiener--Hopf
equations with triangular matrix functions containing exponential
factors. More precisely finding functions \(\Phi_-^{(0)}(\alpha)\),
\(\Phi_-^{(L)} (\alpha)\),
\(\Psi_+^{ (0)}(\alpha)\)
and \(\Psi_+^{ (L)}(\alpha)\)
analytic in respective half-planes, satisfying the relationship
\begin{equation}
  \label{eq:main}
\begin{pmatrix}
 \Phi_-^{(0)}(\alpha) \\
 \Phi_-^{(L)} (\alpha)
 \end{pmatrix}=\begin{pmatrix}
  A(\alpha)&  B(\alpha)e^{i \alpha L} \\
 C(\alpha)e^{-i \alpha L} & 0
 \end{pmatrix}\begin{pmatrix}
 \Psi_+^{ (0)}(\alpha) \\
 \Psi_+^{ (L)}(\alpha) 
 \end{pmatrix}+ \begin{pmatrix}
 f_1(\alpha)\\
f_2 (\alpha)
 \end{pmatrix},
\end{equation}
on the real line. The remaining functions \(A(\alpha)\), 
\(B(\alpha)\) and \(C(\alpha)\) are known and \(L\) is a positive
constant. Many problems have a Wiener--Hopf equation of this type~\citep{Antipov, Aktosun1992} or
could be reduced to such equation by performing matrix manipulations. 
 Presently no complete solution of \eqref{eq:main} is known. We note that there is an additional difficulty in finding a
factorisation of the matrix in \eqref{eq:main} because of the presence
of analytic functions \(e^{i \alpha L}\)
and \(e^{-i \alpha L}\)
which have exponential growth in one of the half-planes. 

In the literature concerning applications of the Wiener--Hopf technique,
one of the most widely used methods is so-called ``pole removal'' or
``singularities matching'' \citep[\S~4.4,~5.3]{Noble};
\citep[\S~4.4.2]{daniele2014wiener}. It has a severe limitation that
certain functions have to be rational or meromorphic. One way to
extend the use of this method is by employing a rational approximation
\citep{Pade, My1}, which was successfully used in
\citep{Abraha_all_pade, Ab_ex, Nigel_cyl}. However, even with
this extension the class of functions which can be solved is rather
limited. In \cite{AYTON_PE} scattering by a finite flat plate with poroelastic extension is considered yielding a 3 x 3 Wiener-Hopf matrix system. The far-field acoustics are obtained by employing rational approximations, however this approach cannot be used to determine mid- or near-field results accurately since the functions involved are not meromorphic and cannot be accurately represented by rational functions everywhere. Additionally, the results were most accurate in \cite{AYTON_PE} when considering high frequency interactions.
In this paper we propose a different extension to the pole
removal technique for functions that have arbitrary
singularities. This new method is suitable for accurately predicting the scattered field everywhere, and is not restrictive on frequency. This new method could also be extended to higher-dimensional matrices, however in this paper we shall specifically only consider 2 x 2 matrices.

The proposed procedure is related to Schwarzschild's series. In the original paper \citep{Schwarzschild1901} Schwarzschild
studied diffraction of a normal incidence plane wave by a slit of finite
length in a perfectly conducting screen. This was achieved by considering the
diffractions from the each half-planes as a sequence of excitations from
the other. This was later extended  to near-normal incidence in \citep{Karp_Russek} and
to all angles in \citep{millar_1958,Grinberg}. Schwarzschild's series relates to a more general framework of Geometric
Theory of Diffraction \citep{Keller:62} and the Physical Theory of
Diffraction, see \citet[\S~8.7.3]{ufimtsev2003theory}. What is common
between all those approaches and the one proposed in this paper is
the idea of solving parts of the problem and then bringing the parts
together in an iterative manner.  Recently, the  Schwarzschild's
series were used in~\citep{Lyu,ROGER2005} but  only  to  obtain  the  scattered  pressure  on  the  surface  of  the  flat  plate,  and  the
far-field  sound  was  obtained  using  the  surface  pressure
integral  based  on  the  theories of  Kirchhoff  and  Curle.

The proposed procedure in this paper could be summarised as follows (see
Section~\ref{sec:2} for details):
\begin{enumerate}
\item A partial factorisation
with exponential factors in the desired half-planes.
\item Additive splitting of some terms.
\item Application of Liouville's theorem.
\item Iterative procedure to determine the remaining unknowns.
\end{enumerate}
Importantly, this method is entirely algorithmic and bypasses the need to construct a multiplicative matrix
factorisation. Theoretical aspect of this method were addressed in~\citep{My_iter}.
We validate the method by considering the two-dimensional Wiener-Hopf problem of the scattering of a sound wave by a finite impermeable plate, which has a known solution \citep{Ellip_book}. We then use the method to investigate the generation of sound by a gust convecting in uniform flow over a semi-infinite impermeable plate with a finite porous extension. We use these results to investigate the effects of a impermeable-permeable junction on the results predicted by \citep{JP}.

We present full details for the
derivation of the porous-extension problem in Section \ref{sec:1},
followed by the iterative Wiener-Hopf method in Section
\ref{sec:2}. In Section \ref{sec:3} we briefly discuss the appropriate formulation of the finite-plate problem. In Section \ref{sec:4a} we present results  validating the new Wiener-Hopf
factorisation method by comparing our finite plate results to the
known solution, then in Section \ref{sec:4b} we present results
for the gust interacting with a porous extension and the effects of an impermeable-permeable junction on the noise reduction predictions of \citet{JP}. Finally we discuss our
conclusions in Section \ref{sec:5}.

\section{Formulation of the problem} 
\label{sec:1}

\begin{figure}
 \centering
 \includegraphics[scale=0.8]{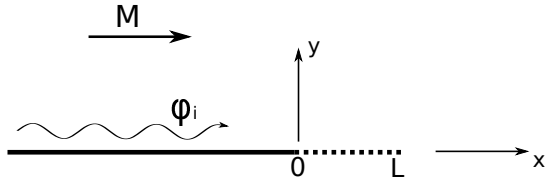}
\caption{Diagram of the model problem: a semi-infinite rigid plate lies in $y=0$, $x<0$, and a finite porous plate lies in $y=0$, $0<x<L$. An unsteady perturbation, $\phi_{i}$, convects with the mean flow in the positive $x$ direction.}
\label{fig:plate}
\end{figure}

In order to investigate the effect of a finite porous edge on trailing-edge
noise, we consider the generation of noise by a convective gust with velocity potential $\phi_{i}(x,y)e^{\i k_{3}z-\i \omega t}$ interacting with a semi-infinite impermeable plate, \(x\in(-\infty,0), y=0\),
with a porous edge, \(x\in(0,L), y=0\),
such that the impermeable-permeable junction is at \(x,y=0\), as illustrated in Figure \ref{fig:plate}. The steady background mean flow is parallel to the plate in the positive $x$-direction and has Mach number $M$.
The matrix Wiener--Hopf problem is obtained by considering the
Fourier transform of appropriate boundary conditions with respect to the ends of the porous plate at
\(x=0\)
and \(x=L\).

We introduce a scattered velocity potential of the form
\[\phi_{1} (x,y)e^{\i k_{3}z-\i\omega t},\]
and in what follows, the time factor
\(e^{-i\omega t}\) will be omitted and we shall neglect the $z$-dependence ($k_{3}=0$).

The  velocity potential \(\phi_i\) associated with a gust incident at angle
\(\theta_i\) to the \(x\) axis with wavenumber $k_{0}$ is
\begin{equation}
  \label{eq:inc}
  \phi_i = \exp(ik_0 x\cos \theta_i-ik_0 y \sin \theta_i ).
\end{equation}

The governing convective Helmholtz equation is
\begin{equation}
 (1-M^{2})\pderiv{^{2}\phi_{1}}{x^{2}}+\pderiv{^{2}\phi_{1}}{y^{2}}-2\i \frac{\omega}{c_{0}}M\pderiv{\phi_{1}}{x}+\frac{\omega^{2}}{c_{0}^{2}}\phi_{1}=0,
\end{equation}
with $\omega=k_{0}\cos\theta_{i} M c_{0}$ and \(c_{0}\) is the speed of sound. 

We apply a convective transform,
\begin{equation}
\phi_{1}=\phi\,\e^{\i k_{0}\cos\theta_{i} M^{2}x/\beta^{2}},
\end{equation}
where $\beta^{2}=1-M^{2}$, and the Prandtl-Glauert transformation ($y\to\beta y$) to reduce the governing equation to
\[\Big(\frac{\partial}{\partial x^2}+\frac{\partial}{\partial y^2}
+w^2 \Big)
\phi=0 ,\]
where $w=\delta M$, with $\delta = k_{0}\cos\theta_{i}/\beta^{2}$. Full details of the convective transform and Prandtl-Glauert scaling can be found in \citet{Tsai}.

The boundary conditions on the impermeable and permeable sections are;
\begin{align}
& \frac{\partial \phi(x,y)}{\partial y}+\frac{\partial \phi_{i}(x,y)}{\partial y}=0,\quad y=0, \quad -\infty<x<0 ,\label{eq:rigidBC}\\
& \frac{\partial \phi(x,y)}{\partial y}+\frac{\partial \phi_{i}(x,y)}{\partial y}= \frac{\mu}{2} \Delta\left(\phi(x,0)+\phi_{i}(x,0)\right), \quad y=0, \quad 0<x<L,\label{eq:porousBC}
\end{align}
respectively, where $\Delta$ stands for the jump across $y=0+$ and $y=0-$, and \(\mu\) is the porosity parameter
($\mu=\alpha_{H}K_{R}/(\upi R^{2}\beta)$) for a porous plate with
evenly-spaced circular apertures of radius $R$, Rayleigh conductivity
of $K_{R}=2R$, and fractional open area  $\alpha_{H}$ (see
\citet{HoweBook}). The conditions on the permeable section are
equivalent to impedance boundary conditions.  We require the length scale of the incident
disturbance to be larger than the aperture radius, $k_{0}R\ll1$, and
the fractional open area to be small, $\alpha_{H}^{2}\ll1$, so that
the porous boundary condition provides a first-order correction
accounting for the perforations in the plate. We shall focus on small
$\mu$ values as this will be beneficial later to ensure rapid
convergence of the iterative method. For comparison, the range of
corresponding $\mu$ values investigated by \citet{JP} in connection to
the silent flight of owl is $\mu\in[10^{-9},12.24]$.

We require
\begin{equation}
\phi(x,0)+\phi_{i}(x,0)=0,  \quad  x>L, \label{eq:wakeBC}
\end{equation}
to ensure continuity of pressure downstream of the plate.
Also \(\frac{\partial \phi(x,y)}{\partial y}\)  (which will be denoted
by \(\phi '(x,y)\) ) must be continuous  across \(y=0\) 
\begin{equation}
\label{eq:sym}
\frac{\partial \phi(x,0+)}{\partial y}=\frac{\partial \phi(x,0-)}{\partial y},  \quad y=0.
\end{equation}
The solution, \(\phi\), is required to satisfies the Sommerfeld
radiation condition for  outgoing waves at infinity,
\begin{equation}
 r^{-1/2} \left( \frac{\partial \phi}{\partial r}-ik_0\phi\right) \to
 0,  \text{  where  } r=\sqrt{x^2+y^2},
\end{equation}
and the edge conditions are taken as in~\citet[\S~2.1]{Noble} to
achieve the least singular solutions.
\begin{subequations}
\begin{align}
&\phi(x,0)\to c_1, \text{  as  } x \to 0-,\\
&\phi(x,0)\to c_2, \text{  as } x \to L-,\\
&\frac{\partial \phi(x,0)}{\partial y} \to c_3x^{-1/2},  \text{ as } x \to 0+,\\
&\frac{\partial \phi(x,0)}{\partial y} \to c_4x^{-1/2}, \text{ as } x \to L+,
\end{align}
\label{eq:edgeconditions}
\end{subequations}
where $c_{i}$ are constants. The two conditions at \(L\) impose the
unsteady Kutta conditions, see~\cite{AYTON_kutta} for more details. 
\subsection{Reduction to Wiener--Hopf Equation}

The next step is to write down the relationship between different
half-range Fourier transforms by using the boundary
conditions. These relations can be combined
to form a matrix Wiener--Hopf equation.

We define the half-range and full-range Fourier transforms with respect to \(x=0\) by
\begin{align*}
\Phi(\alpha, y)=& \int_{-\infty}^{0}\phi( \xi,y) e^{i \alpha \xi} d \xi
+\int^{\infty}_{0}\phi (\xi,y)  e^{i \alpha \xi} d \xi,\\
 =& \Phi^{(0)}_-( \alpha,y)+\Phi^{(0)}_+(\alpha,y),
\end{align*}
and the Fourier transforms with respect to the point \(x=L\) by
\begin{align*}
\Phi^{(L)}( \alpha,y)=& \int_{-\infty}^{L}\phi ( \xi,y) e^{i \alpha (\xi-L)} d \xi
+\int^{\infty}_{L}\phi ( \xi, y)  e^{i \alpha (\xi-L)} d \xi,\\
 =& \Phi^{(L)}_-( \alpha,y)+\Phi_{+}^{(L)}(\alpha, y).
\end{align*}
At the end of the analysis the inverse Fourier transforms are
applied. The relation between the transforms is
\[ \Phi^{(L)} (\alpha,y)=\Phi ( \alpha,y) e^{-i \alpha L}.\]
We can also define the transform of the finite interval only
\begin{displaymath}
\Phi_1(\alpha, y)=\int_{0}^{L}\phi( \xi,y) e^{i \alpha \xi} d \xi,
\end{displaymath}
and note that
\begin{displaymath}
\Phi_1(\alpha, y)=e^{i \alpha L}\Phi^{(L)}_-( \alpha,y) -\Phi^{(0)}_-( \alpha,y).
\end{displaymath}
Now transforming the boundary conditions for the rigid plate, \eqref{eq:rigidBC}, we obtain
\begin{equation}
\Phi^{'(0)}_-( \alpha,0)=-\Phi^{'(0)}_{i\,-}( \alpha,0).\label{eq:rigidFT}
\end{equation}
From the condition on the porous plate, \eqref{eq:porousBC},
\begin{equation}
\Phi_{1}^{'} (\alpha, 0) +\Phi_{i\, 1}^{'} (\alpha, 0)= \mu \Phi_1(\alpha,0).\label{eq:porousFT}
\end{equation}
  where we use \(\Phi_1(\alpha,0+)=-\Phi_1(\alpha,0-)\).
Finally from \eqref{eq:wakeBC} we obtain
 \begin{equation}
 \Phi_{+}^{(L)}(\alpha, 0)=0.\label{eq:wakeFT}
 \end{equation}
The solution to the governing equation can be written as
 \begin{equation}
\Phi(\alpha, y) = \text{sgn}(y)A(\alpha)e^{-\gamma |y|},\label{eq:solFT}
\end{equation}
where $\gamma(\alpha)=\sqrt{\alpha^{2}-w^{2}}$ (were the branch cuts
are chosen in the standard way, see~\citet[pg 9]{Noble}.
The Wiener-Hopf
equations are now obtained from the following relationships 
\begin{align}
& \Phi^{'(0)}_{-}+\Phi_1^{'}+e^{i \alpha L}\Phi^{'(L)}_{+}=-\gamma e^{i \alpha L}(\Phi^{(L)}_++\Phi^{(L)}_-),\label{eq:1} \\
& \Phi^{'(0)}_-+\Phi^{'(0)}_+=-\gamma e^{i \alpha L}(\Phi^{(L)}_++\Phi^{(L)}_-),\label{eq:2}
\end{align}
by substituting the known transforms \eqref{eq:rigidFT},\eqref{eq:porousFT} and \eqref{eq:wakeFT}, and using the relationship \(\Phi'(\alpha, y)=-\gamma \Phi(\alpha, y)\). 
The Wiener--Hopf equations can now finally be written (by substituting an
expression for \(\Phi^{(L)}_-\) from \eqref{eq:1} to \eqref{eq:2}) as:
\begin{equation}
\begin{pmatrix}
\Phi_-^{(0)}(\alpha) \\
\Phi_-^{(L)} (\alpha)
\end{pmatrix}=-\begin{pmatrix}
 \frac{1}{\gamma(\alpha)}-P  &  Pe^{i \alpha L} \\
\frac{1}{\gamma(\alpha)}e^{-i \alpha L} & 0
\end{pmatrix}\begin{pmatrix}
\Phi_+^{ '(0)}(\alpha) \\
\Phi_+^{ '(L)}(\alpha) 
\end{pmatrix}- \begin{pmatrix}
f_1(\alpha)\\
f_2 (\alpha)
\end{pmatrix},
\label{eq:main_mat}
\end{equation}
where \(P=\frac{-1}{\mu}\). The forcing provided by the gust is
\begin{equation}
f_1=\frac{k_0\sin \theta_i}{\gamma(\alpha)(\alpha+\delta)} +\frac{k_0\sin \theta_i}{(\alpha+\delta)}(-P+Pe^{i(\alpha+\delta)L})+\frac{1}{i(\alpha+\delta)}e^{i(\alpha+\delta) L},\label{eq:f1}
\end{equation}
and
\begin{equation}
f_2=\frac{k_0\sin \theta_i}{\gamma(\alpha)(\alpha+\delta)} e^{-i \alpha L}+\frac{1}{i(\alpha+\delta)}e^{i\delta L}. \label{eq:f2}
\end{equation}
We note that on the left hand side of \eqref{eq:main_mat} the unknown functions are minus
half-transforms, \(\Phi_-^{(0)}(\alpha)\)
and \(\Phi_-^{(L)} (\alpha)\),
and on the right are the derivatives of the plus half-transforms, \(\Phi_+^{'(0)}(\alpha)\)
and \(\Phi_+^{'(L)} (\alpha)\).

The behaviour of various transforms at infinity is determined by the edge conditions \eqref{eq:edgeconditions}.  We have \(\Phi_-^{(0)}(\alpha) \to \alpha^{-1}\) and
\(\Phi_-^{(L)}(\alpha) \to \alpha^{-1}\)   as \(\alpha \to
\infty\) in the lower half-plane, and  \(\Phi_+^{
'(0)}(\alpha)\to \alpha^{-1/2}\) and \(\Phi_+^{ '(L)}(\alpha)\to
\alpha^{-1/2}\) as \(\alpha \to \infty\) in the upper half-plane.

\subsection{Far-field acoustic directivity}
Supposing we can solve \eqref{eq:main_mat} for the unknown $\Phi$ terms, we can determine the full acoustic solution, \eqref{eq:solFT}, which can be written as
\begin{equation*}
 \Phi(\alpha,y)=\Phi_{-}^{(L)}(\alpha,0)\text{sgn}(y)e^{-\gamma|y|}
\end{equation*}
This can be inverted to obtain the velocity potential
\begin{equation*}
 \phi(x,y)=\frac{\text{sgn}(y)}{2\upi}\int_{-\infty}^{\infty}\Phi_{-}^{(L)}(\alpha,0)e^{-\i\alpha x-\gamma|y|}d\alpha,
\end{equation*}
which for $x,y\to\infty$ can be approximated by the method of steepest descents to yield
\begin{equation*}
 \phi(r,\theta)\sim \frac{\sqrt{w}e^{-\upi i/4}}{\sqrt{2\upi}}\Phi_{-}^{(L)}(-w\cos\theta,0)\sin\theta\frac{e^{i wr}}{\sqrt{r}}
\end{equation*}
where $(r,\theta)$ are polar coordinates measured from the rigid-porous junction, $x=y=0$.
The far-field directivity, $D(\theta)$, is then defined via
\begin{equation}
 \phi(r,\theta)\sim D(\theta)\frac{e^{i wr}}{\sqrt{r}}, \qquad r\to\infty.\label{eq:directivity}
\end{equation}

\section{Approximate  Wiener--Hopf Factorisation}
\label{sec:2}

We look for an approximate matrix factorisation of
\eqref{eq:main_mat}. We will be using the standard results in the
theory of scalar Wiener--Hopf factorisation, and in what follows a superscript/subscript  \(+\) (or \(-\)) indicates  that the
function is \emph{analytic  in the upper (lower)
  half-plane}. Functions without the superscript/subscript are known. 
A convention is introduced here to distinguish the additive and the
multiplicative factorisation by using a superscript and subscript
notation (e.g \(K(\alpha)=K^-(\alpha)+K^+(\alpha)\) for additive
Wiener--Hopf splitting and  \(K(\alpha)=
K_-(\alpha)K_+(\alpha)\) for multiplicative factorisation). For more details on the
scalar factorisation see \citep{Noble, daniele2014wiener}.

The most characteristic aspect of the Wiener--Hopf method is the
application of Liouville's theorem in order to obtain two separate
connected equations from one equation.  In order for Liouville's
theorem to be used two conditions have to be satisfied: the
analyticity and (at most) polynomial growth at infinity.  These two
conditions will be treated here in turn. First, a partial factorisation is
considered that has the exponential functions in the right place (in
the half-planes where they decay exponentially) and
some of the required analyticity. To do so we premultiply \eqref{eq:main_mat} by a suitably chosen matrix yielding
\begin{align*}
 \lefteqn{\begin{pmatrix}
\frac{-e^{-i \alpha L} }{P_-} &\frac{1-\gamma P }{P_-}\\
 \frac{1 }{P_-} & 0
 \end{pmatrix}\begin{pmatrix}
  \Phi_-^{(0)}\\
\Phi_-^{(L)} 
 \end{pmatrix}} \\ 
&\hspace*{2cm}=-\begin{pmatrix}
  0  &   - P_+ \\
\frac{ P_+}{\gamma P } - P_+&   P_+e^{i \alpha L}
 \end{pmatrix}\begin{pmatrix}
  \Phi_+^{ '(0)} \\
 \Phi_+^{ '(L)} 
 \end{pmatrix}+ \begin{pmatrix}
 f_3\\
f_4  \end{pmatrix},
\end{align*}
or re-arranged 
\begin{align*}
 \lefteqn{\begin{pmatrix}
\frac{-e^{-i \alpha L} }{P_-} &\frac{1 }{P_-}\\
 \frac{1 }{P_-} & 0
 \end{pmatrix}\begin{pmatrix}
 \Phi_-^{(0)} \\
\Phi_-^{(L)}
 \end{pmatrix}+\begin{pmatrix}
- \gamma P_+\Phi_-^{(L)} \\
  \frac{1}{\gamma P_-} \Phi_+^{ '(0)}
 \end{pmatrix}}\\ 
&\hspace*{2cm}=-\begin{pmatrix}
  0  &   - P_+ \\
 - P_+&  P_+ e^{i \alpha L}
 \end{pmatrix}\begin{pmatrix}
 \Phi_+^{ '(0)} \\
\Phi_+^{ '(L)} 
 \end{pmatrix}+
\begin{pmatrix}
 f_3\\
f_4 
\end{pmatrix}
,
\end{align*}
where \(f_4=-f_1/P_-\) (with $f_{1}$ given by \eqref{eq:f1}) and \(f_3\) defined as
\begin{displaymath}
f_3=\frac{k_0\sin \theta_ie^{-i\alpha L}P_+}{(\alpha+\delta)} [ 2e^{i(\alpha+\delta)L}]+\frac{1}{i(\alpha+\delta)}e^{i\delta L}\gamma P_{+}.
\end{displaymath}
Even though our example problem of a finite porous section has
$P=\text{const}$, this method extends to $P(\alpha)$ with at most
polynomial growth at infinity,  for example in the case of an elastic
edge $P(\alpha)$ would be a polynomial of degree \(4\). Therefore we include the relevant details of the factorisation in the more general case, which is suitable for tackling any matrix of the form \eqref{eq:main}.

The next step is to perform some additive splittings;
\begin{align*}
&\gamma P_+\Phi_-^{(L)}= (\gamma P_+\Phi_-^{(L)})^-+(\gamma P_+\Phi_-^{(L)})^+,\\
& \frac{1}{\gamma P_-} \Phi_+^{ '(0)}=(\frac{1}{\gamma P_-} \Phi_+^{ '(0)})^-+(\frac{1}{\gamma P_-} \Phi_+^{ '(0)})^+.
\end{align*}
Note that we cannot calculate these directly since \(\Phi_-^{(L)}\) and
\(\Phi_+^{ '(0)}\) are unknown. An approximation for the splittings
will be found later. Note the similarities with the ``pole removal'' or
``singularities matching'' \citet[\S~4.4,~5.3]{Noble}
\citet[\S~4.4.2]{daniele2014wiener} where one of \(()^-\) or \(()^+\)
would be an infinite sum of unknown coefficients. Those unknown
coefficients are determined later in an approximate fashion. 

We would
like to note that \eqref{eq:main_mat} cannot be fully solved by first
performing rational approximation and then  ``pole removal'' method. The
reason is that a rational approximation of \(\gamma\) or
\(\frac{1}{\gamma}\) which is accurate on the whole real line is not
possible. This is due to the fact that \(\gamma \to |\alpha| \) as \(
|\alpha| \to \infty \) and no rational function has this property. It
is on the other hand possible to obtain good rational approximation on
an interval of the real line and
that in turn could be used to obtain far-field solution using steepest
descent method. So in this
case the proposed method is the only way of fully extending ``pole removal'' technique.

Now  Liouville's theorem can be applied
because the exponential functions are in the correct place and all the
functions have the desired analyticity. Thus, we can apply the
Wiener--Hopf procedure as usual and the four equations become
\begin{align*}
\frac{-e^{-i \alpha L} }{P_-} \Phi_-^{(0)}+\frac{1
}{P_-}\Phi_-^{(L)}-\left(\gamma P_+\Phi_-^{(L)}\right)^-- f_3^-=& 0,\\
P_+\Phi_+^{ '(L)}+\left(\gamma P_+\Phi_-^{(L)}\right)^++ f_3^+=&0,\\
 \frac{1 }{P_-} \Phi_-^{(0)}+
\left(\frac{1}{\gamma P_-} \Phi_+^{ '(0)}\right)^--f_4^-=& C,\\
 P_+ \Phi_+^{ '(0)}-P_+ e^{i \alpha L}\Phi_+^{ '(L)} -\left(\frac{1}{\gamma P_-} \Phi_+^{ '(0)}\right)^++f_4^+=&C,
\end{align*}
where \(C\) is a constant. It turns out that this constant plays no
role as is discussed at the end of Section \ref{sec:3a}, so can be
taken to be zero. 
The four equations can be rearranged (and their order changed):
\begin{align}
\Phi_-^{(L)}=& P_-\big(\left(\gamma P_+\Phi_-^{(L)}\right)^-+ f_3^-\big) +e^{-i
  \alpha L} \Phi_-^{(0)}, \label{eq:eq3} \\ 
\Phi_+^{ '(L)}=&\frac{1}{P_+}\big(-\left(\gamma P_+\Phi_-^{(L)}\right)^+-
f_3^+\big) \label{eq:eq1}, \\
\Phi_+^{ '(0)}=&  \frac{1}{P_+}\big(\left(\frac{1}{\gamma P_-} \Phi_+^{
   '(0)}\right)^+-f_4^+\big)+e^{i \alpha L}\Phi_+^{ '(L)}. \label{eq:eq4}\\
 \Phi_-^{(0)}=&P_-\big(-\left(\frac{1}{\gamma P_-} \Phi_+^{
   '(0)}\right)^-+f_4^-\big),\label{eq:eq2}
\end{align}

 When the equations are written in this form  it is clear that
if \(\Phi_-^{(L)}\) is known then it could be used to calculate  
\(\Phi_+^{ '(L)}\) from  \eqref{eq:eq1} and this, in turn, produces
\(\Phi_+^{ '(0)}\) by looking at \eqref{eq:eq4} followed
by the calculation of
\(\Phi_-^{(0)}\) in \eqref{eq:eq2} and then it loops round. This
observation will form the basis to the iterative procedure.  So far the solution is exact but in order to make progress an
approximation is needed.

In order to solve approximately we will describe an iterative procedure and denote the \(n\)th iteration by \(\Phi_-^{(L)n}\). The equation \eqref{eq:eq3} is
 going to be approximated by  neglecting the \(e^{-i
  \alpha L} \Phi_-^{(0)}\) term (which as we will see later is small
in many cases) i.e.
\begin{equation}
\label{eq:sca1}
  \Phi_-^{(L)1}=P_-\big((\gamma P_+\Phi_-^{(L)1})^-+ f_3^-\big), 
 \end{equation}
which corresponds to only considering  the junction at \(x=L\).  In
other words we are considering the problem of scattering from
a semi-infinite porous plane. Note that \(\Phi_-^{(L)1}\) is an approximation to \(\Phi_-^{(L)}\)
but importantly \(\Phi_-^{(L)1}\) have the same analyticity properties as
\(\Phi_-^{(L)}\). The above equation \eqref{eq:sca1} can be rearranged as a scalar Wiener--Hopf equation in the following manner,
\begin{equation*}
  \left(\frac{1-\gamma P }{P_-}\Phi_-^{(L)1}\right)^-=f_3^-.
 \end{equation*}
We introduce an unknown function \(D_+\) defined by
\begin{equation*}
  \left(\frac{1-\gamma P }{P_-}\Phi_-^{(L)1}\right)^+= D^+,
 \end{equation*}
then adding together yields
\begin{equation}
\label{eq:W-H1}
\frac{1-\gamma P }{P_-}\Phi_-^{(L)1}= D^++ f_3^-,
 \end{equation}
which is a Wiener--Hopf equation, with solution
\begin{equation*}
\Phi_-^{(L)1}=\frac{P_-}{(1-\gamma P)_-}\left(\left( \frac{ f_3^-}{ (1-\gamma P)_+ }\right)^-  \right) .
\end{equation*}
Also we are able to compute \(\left(\gamma P_+\Phi_-^{(L)1}\right)^+\)  (which we need in \eqref{eq:eq1}) by considering \(D^+\). Note that
\begin{equation*}
  D^+ =\left(\frac{1-\gamma P }{P_-}\Phi_-^{(L)1}\right)^+=-\left(\gamma P_+\Phi_-^{(L)1}\right)^+,
 \end{equation*}
 and solving the Wiener--Hopf equation  \eqref{eq:W-H1} for \(D^+\) gives
 \begin{equation*}
  D^+ =  (1-\gamma P)_+\left( -\left( \frac{ f_3^-}{ (1-\gamma P)_+ }\right)^+ \right) .
 \end{equation*}
So we are solving this Wiener--Hopf matrix factorisation by
first considering the scalar equation which arises from one of the
junctions (\(x=L\)) in the boundary conditions. 
 So now using \eqref{eq:eq1} we easily obtain an expression for
 \(\Phi_+^{ '(L)1}\). This in turn is used in \eqref{eq:eq4} to reduce
 the solution for \(\Phi_+^{ '(0)1}\) to a scalar Wiener--Hopf problem
 in the same fashion as \eqref{eq:sca1}. This is now coupling junction
 at \(x=L\) with the junction at \(x=0\), and gives
 \begin{equation*}
\Phi_+^{ '(0)1}=\frac{\gamma_+}{(\gamma P-1)_+}\left(\left(\frac{(-f_4^++P_+e^{i \alpha L}\Phi_+^{ '(L)1})P_-\gamma_-}{(\gamma P-1)_-}\right)^+\right).
\end{equation*}
 
  An expression for  \(\Phi_-^{ (0)1}\) can in turn be obtained from \eqref{eq:eq2}. In order to simplify again note that 
   \begin{equation*}
 \left(\frac{1}{\gamma P_-} \Phi_+^{
   '(0)1}\right)^-=\frac{(\gamma
 P-1)_-}{P_-\gamma_-}\left(-\left(\frac{(-f_4^++P_+e^{i \alpha
       L}\Phi_+^{ '(L)1})P_-\gamma_-}{(\gamma P-1)_-}\right)^-\right).
   \end{equation*}
  Once \(\Phi_-^{(0)1}\) is computed this can be looped round to
  calculate \(\Phi_-^{(L)2}\) from using the full equation
  \eqref{eq:eq3} and \eqref{eq:eq2}. As the loops are performed the
  coupling between the junction \(x=L\) and \(x=0\) is successively corrected and  \(\Phi_+^{' (0)n}\) approaches the correct solution. As it is shown in the next sections, in
  practice in most case the convergence is very fast and the desired accuracy is
  achieved for \(n=2\) or \(n=3\).   

The formula, after some computation, for the iterations would be 
\begin{equation}
  \label{eq:iteration-psi}
\Phi_+^{ '(0)n}= \frac{-1}{(K_{3})_+}\left(\frac{f_4^++e^{i \alpha
      L}\left( \left( \gamma P_+\Phi_-^{(L)n-1}\right)^++f_3^+\right)}{(K_3)_-}\right)^+,
\end{equation}
where \(K_3=\frac{\gamma
  P-1}{\gamma P_-}\) and also
\begin{equation}
\label{eq:iteration-phi}
\Phi_-^{(L)n}=\frac{1}{(K_{1})_-}\left(\frac{f_3^- +e^{-i
  \alpha L}\left(-\left( \frac{1}{\gamma P_-}  \Phi_+^{'(0)n}\right)^-
+f_4^-\right) }{(K_1)_+}\right)^-,
\end{equation}
 with \(K_1=\frac{\gamma P-1 }{P_-}\). 

\subsection{Convergence}
\label{sec:Con}

The convergence of the proposed procedure was considered
in~\citep{My_iter}. It gave some of the sufficient conditions for
convergence but not the necessary conditions. In this paper the class of matrix~\eqref{eq:main}
is more general than is considered in that paper. Two main differences
are that: we have no common strip of analyticity in ~\eqref{eq:main}
and \(A\), \(B\) and \(C\) do not belong to the classes described
in~\citep{My_iter}. With respect to the later, on the application of Liouville's
theorem the entire functions are constant rather than zero, this is discussed in
Section \ref{sec:3a}.

In the case when \(k_0\) has a small imaginary part the convergence
analysis follows almost unchanged as in~\citep{My_iter}. This gives
that convergence occurs for large enough \(|k_0|\), and is faster the
larger \(|k_0|\) is. We found that the procedure converges fast, for a large range of parameters, even without ``the
strip of analyticity'' in other words we have taken \(k_0\) without an
imaginary part. This is unlike 
other methods such as rational approximation and ``pole removal'' rely
on taking non-zero imaginary part in \(k_0\). In this paper we also found that the
results converge faster for large \(|k_0|\) (which is equivalent to
larger \(L\)) as was is found in
~\citep{My_iter}.

Note that the issue of
convergence and the issue of having good first approximation \eqref{eq:sca1} are
quite different. It is very much possible to have a very bad first
approximation which converges fast to the correct solution (see Figure
5 in~\citep{My_iter}). On
the other hand it is also possible to have a situation where no matter
how close  to the exact solution is the
initial guess (as long as it is not exactly the solution) each iteration will make the solution only further away
from the exact solution, so diverges. Of course in the case where there
is convergence picking a better first guess means that the error in
the iterations is smaller and so fewer iterations are needed.
A better first guess can be achieved sometimes  by first considering \eqref{eq:eq4} rather than
\eqref{eq:eq3} (and neglecting the exponential term) or by keeping some of
the exponential term e.g. \(e^{-i
  \alpha L}P_-f_4^-\) in \eqref{eq:eq3}.

For the acoustic problems considered here we have found that the
second and the third iteration were very close together, in almost all
cases. This means
that \(\Phi_-^{(L)3}\) and \(\Phi_+^{ '(0)3}\)
satisfy~\eqref{eq:main_mat} almost exactly and hence are very close to the
exact solution. In summary, although we could not guarantee \emph{a
  priori} that the method would converge for~\eqref{eq:main_mat}, once
it is found to converge we can deduce it is close to the correct solution.  

\section{A finite plate}
\label{sec:3}

\subsection{Wiener--Hopf formulation}
\label{sec:3a}
As a way of verifying the procedure in the previous section we consider a simpler problem to
which an exact solution can be constructed using Mathieu functions in
the limit of zero porosity. Even for this simple
problem the resulting Wiener--Hopf matrix equation cannot be solved
exactly.

We look at the situation where the semi-infinite rigid plate is
not present, only the finite porous plate. We also suppose there is no background flow, and the incident perturbation is given by
\begin{equation*}
  \phi_i = \exp(-ik_0 x\cos \theta_i-ik_0 y \sin \theta_i-\i k_{0}c t ).
\end{equation*}
The governing equation for the scattered field is 
\[\Big(\frac{\partial}{\partial x^2}+\frac{\partial}{\partial y^2}
+k_{0}^2 \Big)
\phi=0 ,\]

The boundary condition required on the plate transforms to
\begin{displaymath}
\Phi_1^{'} (\alpha, 0) = \mu \Phi_1(\alpha,0)-\frac{k_0 \sin \theta_i}{\alpha-k_0 \cos \theta_i} \left( e^{i(\alpha-k_0\cos\theta_i)L}-1\right),
\end{displaymath}
 and since there is no plate for $x<0$ and $x>L$ we require
 \begin{displaymath}
 \Phi_{+}^{(L)}(\alpha, 0)=0=\Phi_{-}^{(0)}(\alpha, 0).
\end{displaymath}
Now the Wiener--Hopf equations arise from the following
\begin{align*}
& \Phi^{'(0)}_++\Phi_1^{'}+e^{i \alpha L}\Phi^{'(L)}_-=-\gamma e^{i \alpha L}(\Phi^{(L)}_++\Phi^{(L)}_-), \\
& \Phi^{(0)}_-+\Phi^{(0)}_+=e^{i \alpha L}(\Phi^{(L)}_++\Phi^{(L)}_-),
\end{align*}
giving
\[\begin{pmatrix}
\Phi_-^{'(0)}(\alpha) \\
\Phi_-^{(L)} (\alpha)
\end{pmatrix}=-\begin{pmatrix}
 -\frac{1}{P} +\gamma(\alpha) &  e^{i \alpha L} \\
-e^{-i \alpha L} & 0
\end{pmatrix}\begin{pmatrix}
\Phi_+^{ (0)}(\alpha) \\
\Phi_+^{ '(L)}(\alpha) 
\end{pmatrix}+ \begin{pmatrix}
g_1(\alpha)\\
0
\end{pmatrix},\]
with
 \begin{displaymath}
g_1(\alpha)=\frac{k_0 \sin \theta_i}{\alpha-k_0 \cos \theta_i}
\left( e^{i(\alpha-k_0\cos\theta_i)L}-1\right).
\end{displaymath}
Again we need to know the growth at infinity of
\(\Phi_-^{'(0)}(\alpha)\), \(\Phi_-^{(L)} (\alpha)\), \(\Phi_+^{
  (0)}(\alpha)\) and \(\Phi_+^{ '(L)}(\alpha)\) in  relevant
half-planes. We have \(\Phi_-^{'(0)}(\alpha) \to \alpha^{-1/2}\) and
\(\Phi_-^{(L)}(\alpha) \to \alpha^{-1}\)   as \(\alpha \to
\infty\) in the lower half-plane and  \(\Phi_+^{
(0)}(\alpha)\to \alpha^{-1}\) and \(\Phi_+^{ '(L)}(\alpha)\to
\alpha^{-1/2}\) as \(\alpha \to \infty\) in the upper half-plane.

For a finite impermeable plate we can set $\mu=0$ (equivalently $P\to-\infty$). In this case an exact solution for the scattering of a plane wave by a plate can be found in terms of Mathieu functions (discussed in the following subsection).

The equation has the same structure as before and is solved using the
same method which we will not repeat here but simply state the
relevant formula. The four equations this time are
\begin{align*}
-e^{-i \alpha L}\Phi_-^{'(0)}-\gamma^-\Phi_-^{(L)}-\left(\gamma^+\Phi_-^{(L)}\right)^-- g_2^-=& C_1,\\
\Phi_+^{ '(L)}+\left(\gamma^+ \Phi_-^{(L)}\right)^++ g_2^+=&C_1,\\
 \Phi_-^{'(0)}+\left(\gamma^- \Phi_+^{ (0)}\right)^--g_1^-=& C_2,\\
 -\gamma^+\Phi_+^{ (0)}- e^{i \alpha L}\Phi_+^{ '(L)} -\left(\gamma^- \Phi_+^{ (0)}\right)^++g_1^+=&C_2,
\end{align*}
where \(g_2= -e^{-i \alpha L}g_1\). For the initial step we have
\begin{equation*}
\Phi_-^{(L)1}= \frac{-1}{(\gamma)_-}\left(\frac{g_4^-}{(\gamma)_+}\right)^-,
\end{equation*}
where we defined \(g_4^-=g_2^-+C_1\), \(g_4^+=g_2^+-C_1\) and \(g_3^-=g_1+C_2\), \(g_3^+=g_1^+-C_2\).
The iterative solution is obtained 


\begin{equation}
\label{eq:iteration-phi}
\Phi_-^{(L)n}= \frac{-1}{(\gamma)_-}\left(\frac{g_4^- +e^{-i
  \alpha L}\left(\left(  \gamma \Phi_+^{ (0)n}\right)^-+g_3^-\right) }{(\gamma)_+}\right)^-,
\end{equation}
and 
\begin{equation}
  \label{eq:iteration-psi}
\Phi_+^{ (0)n}= \frac{1}{(\gamma)_+}\left(\frac{g_3^++e^{i \alpha L}\left( -\left( \gamma \Phi_-^{(L)n-1}\right)^++g_4^+\right)}{(\gamma)_-}\right)^+.
\end{equation}

We notice from the formulation above that the constants $C_{1,2}$ cancel at each stage. For example,
\begin{equation}
 \Phi_+^{ (0)1}=\frac{1}{(\gamma)_+}\left(\frac{g_3^++e^{i \alpha L}\left( \left( \gamma_{+} \left(\frac{g_4^-}{(\gamma)_+}\right)^-\right)^++g_4^+\right)}{(\gamma)_-}\right)^+,\label{eq:phiconsts}
\end{equation}
where 
\begin{equation*}
  \left( \gamma_{+} \left(\frac{g_4^-}{(\gamma)_+}\right)^-\right)^+= \left( \gamma_{+} \frac{g_4^-}{(\gamma)_+}-\gamma_{+}\left(\frac{g_4^-}{(\gamma)_+}\right)^+\right)^+=-\gamma_{+}\left(\frac{g_4^-}{(\gamma)_+}\right)^+.
\end{equation*}
Considering only the terms in \eqref{eq:phiconsts} that contain constants $C_{1,2}$ we find
\begin{equation*}
 \frac{1}{(\gamma)_+}\left(\frac{-C_{2}-e^{i \alpha L}\left(\gamma_{+}\left(\frac{C_{1}}{\gamma_{+}}\right)^{+}-C_{1}\right)}{\gamma_{-}}\right)^{+}=0,
\end{equation*}
hence neither $C_{1}$ nor $C_{2}$ contribute towards the calculation of $\Phi_{+}^{(0)1}$. A similar calculation shows neither contribute to $\Phi_-^{(L)2}$, and so on for all iterative steps.

\subsection{Exact solution}

An exact solution of acoustic scattering by a finite rigid plate
currently cannot be obtained using a Wiener--Hopf formulation but it is
possible using other methods. It is achieved by considering the Helmholtz
equation in an elliptic coordinate system~\citep{Ellip_book}. The advantage of this coordinate transformation is that the boundary condition on a finite-range interval, i.e. a degenerate ellipse, maps to a full-range elliptic coordinate interval. 

Here we present the exact solution without derivation (which can be
found in \citep{Ellip_book}). To transform from the Cartesian coordinate to the elliptic coordinates
we use the formula
\begin{equation}
x=\frac{d}{2}\cosh \xi \cos \eta \quad y=\frac{d}{2}\sinh \xi \sin \eta
\end{equation}
where \(d\) is the focal distance of the ellipses. 

 We first need to introduce the even 
and odd angular Mathieu functions, which are denoted by
\[ H_m (\eta;q)=
  \begin{cases}
     ce_m(\eta;q),     & \quad  m=0, 1,2 \dots \\
     se_m(\eta;q),  & \quad m= 1,2, 3 \dots\\
  \end{cases}
\]

The even and odd radial Mathieu functions of first and second order
are denoted by
\[ G_m(\xi;q) =
  \begin{cases}
     Je_m(\xi;q),  Jo_m(\xi;q),   & \quad  \text{first kind}, \\
     Ne_m(\xi;q),  No_m(\xi;q), & \quad \text{second kind},\\
  \end{cases}
\]

and finally the Mathieu-Hankel function is given by
\begin{equation*}
Ho^{(1)}_m(\xi;q)=Jo_m(\xi;q)+iNo_m(\xi;q).
\end{equation*}

The scattered acoustic field can be thus expressed as in \citet[19.14]{Ellip_book}
\begin{equation*}
\sum_{r=0}^{\infty} S_rHo_r(\xi)se_r(\eta),
\end{equation*}
where the coefficients \(S_r\) are found to be 
\begin{equation*}
S_n=-i^n\sqrt{8\pi} se_n(\alpha)\frac{Jo'_n(0)}{ Ho'_n(0)}.
\end{equation*}
This solution is valid everywhere. In order to calculate the
far-field numerically an asymptotic formula is used for \(Ho_r(\xi)\).
We shall refer to this solution as the `Mathieu solution'.

\section{Results}\label{sec:4}
Here we investigate the effects of a finite porous extension on the generation of noise by of a gust convecting past a trailing edge by implementing the new Wiener-Hopf factorisation procedure as outlined in the preceding sections. First we illustrate the convergence of the new factorisation method by considering the scattering of a plane sound wave by a finite rigid plate in zero mean flow, for which the solution is known. 

\subsection{Validation of the Iterative Method}\label{sec:4a}

We begin by illustrating our new iterative factorisation method by considering the scattering of a plane sound wave of frequency $k_{0}$ and incidence angle $\theta_{i}$ by a finite impermeable plate of length $2$ as discussed in Section \ref{sec:3}.
The iterative procedure converges quickly when the initially neglected term, $e^{-i\alpha L}\Phi_{-}^{(0)}$ is much smaller than the terms initially retained \citep{My_iter}. It is known from \citet{AytonPeake2013} that for high-frequency finite-plate acoustic interactions, the dominant contribution to the scattered field can be calculated from only one edge (as if the plate were semi-infinite). The correction term due to the rescattering of this acoustic field by the second edge is $O(k_{0}^{-1/2})$ smaller, except at shallow angles directly upstream and downstream where the two fields are formally of the same magnitude to enforce edge continuity conditions. Subsequent correction terms continue to be $O(k_{0}^{-1/2})$ smaller than the last term calculated.

\begin{figure}
 \centering
\begin{subfigure}[b]{0.45\linewidth}
 \centering
 \includegraphics[scale=0.65]{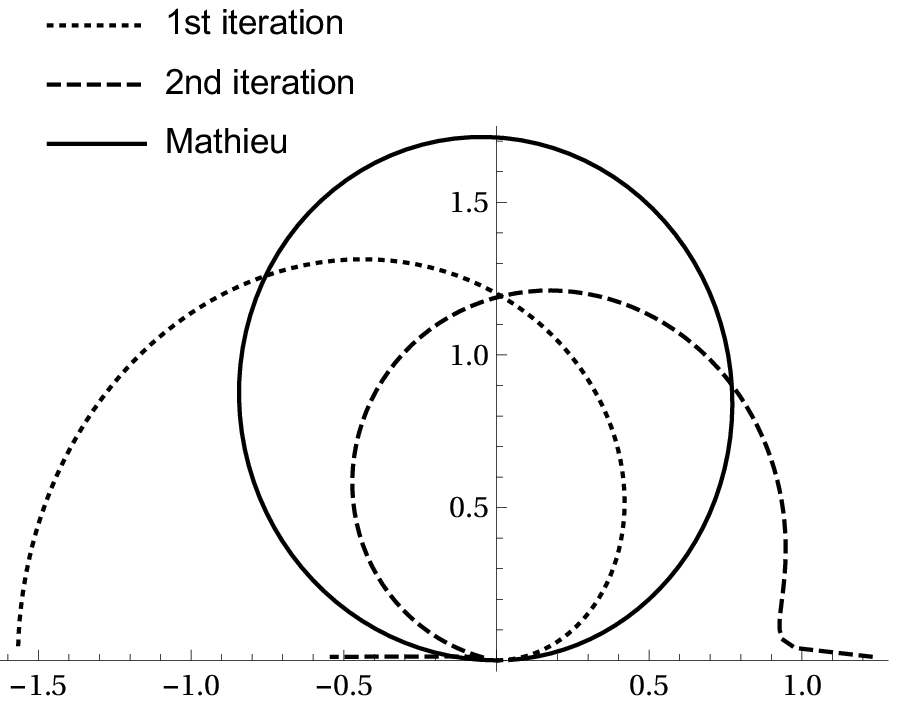}
\caption{$k_{0}=1$.\vspace{5pt}}
\label{fig:finite1}
\end{subfigure}
\hfill
\begin{subfigure}[b]{0.5\linewidth}
 \centering
 \includegraphics[scale=0.65]{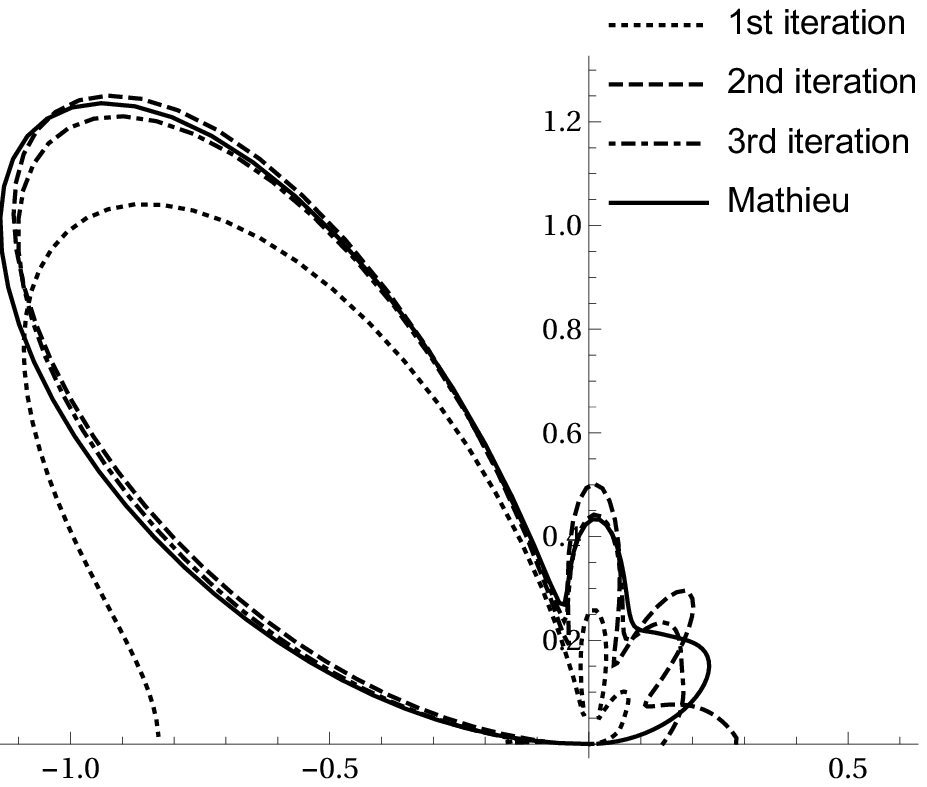}
\caption{$k_{0}=6$.\vspace{5pt}}
\end{subfigure}
\hfill
\begin{subfigure}[t]{0.45\linewidth}
 \centering
 \includegraphics[scale=0.65]{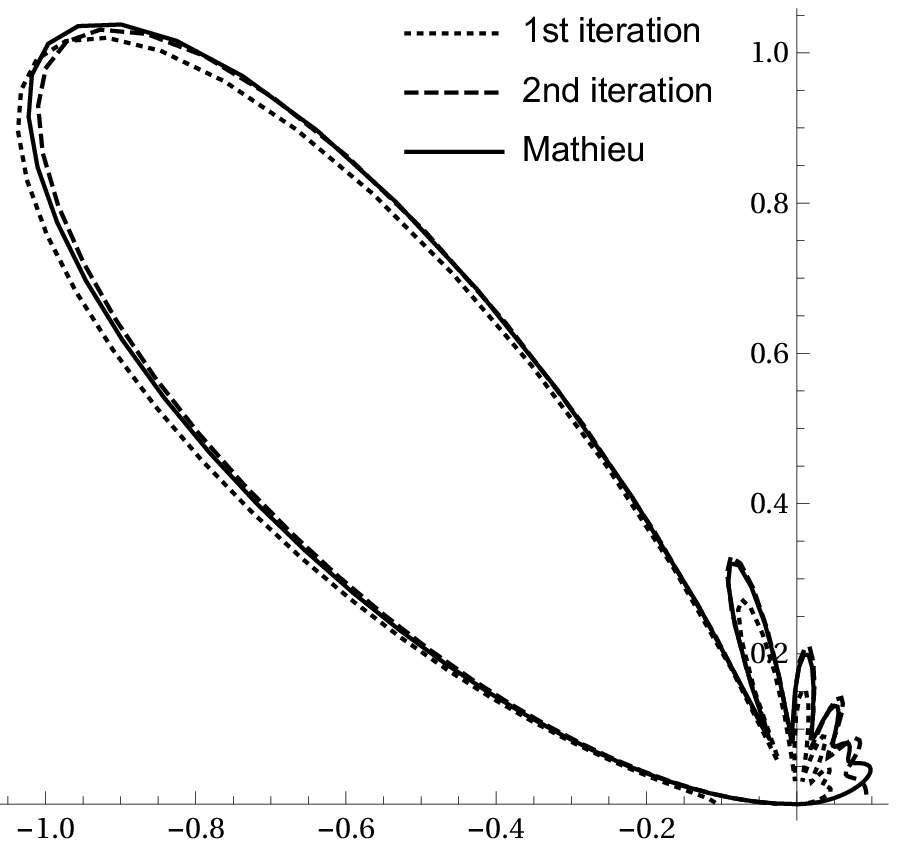}
\caption{$k_{0}=10$.\vspace{5pt}}
\end{subfigure}
\hfill
\begin{subfigure}[t]{0.5\linewidth}
 \centering
 \includegraphics[scale=0.65]{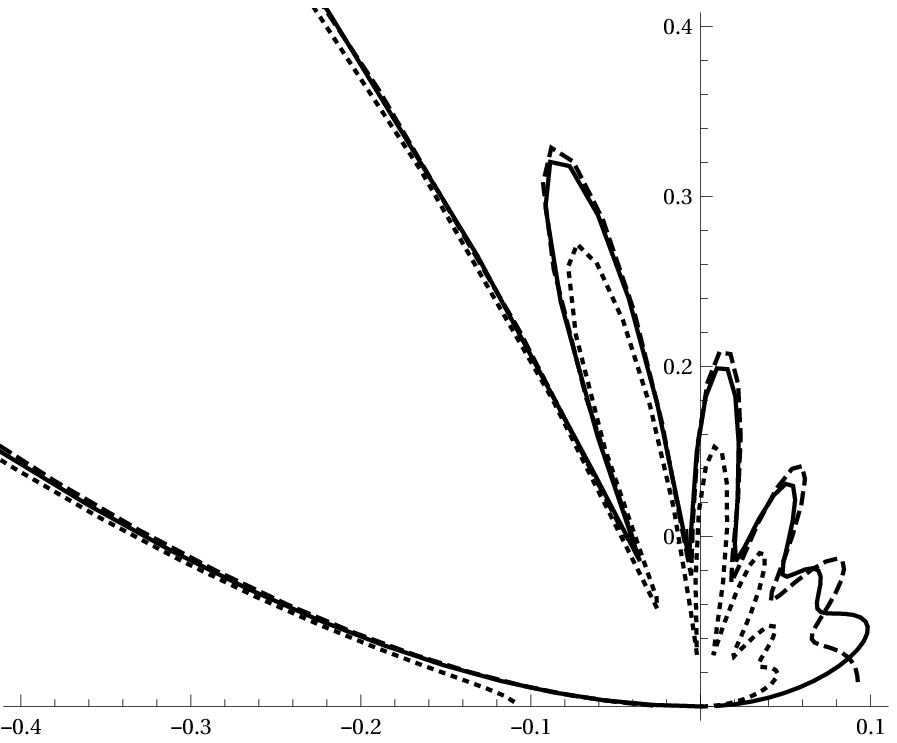}
\caption{$k_{0}=10$ view of highly oscillatory region.}
\end{subfigure}
\hfill
\begin{subfigure}[t]{0.45\linewidth}
 \centering
 \includegraphics[scale=0.65]{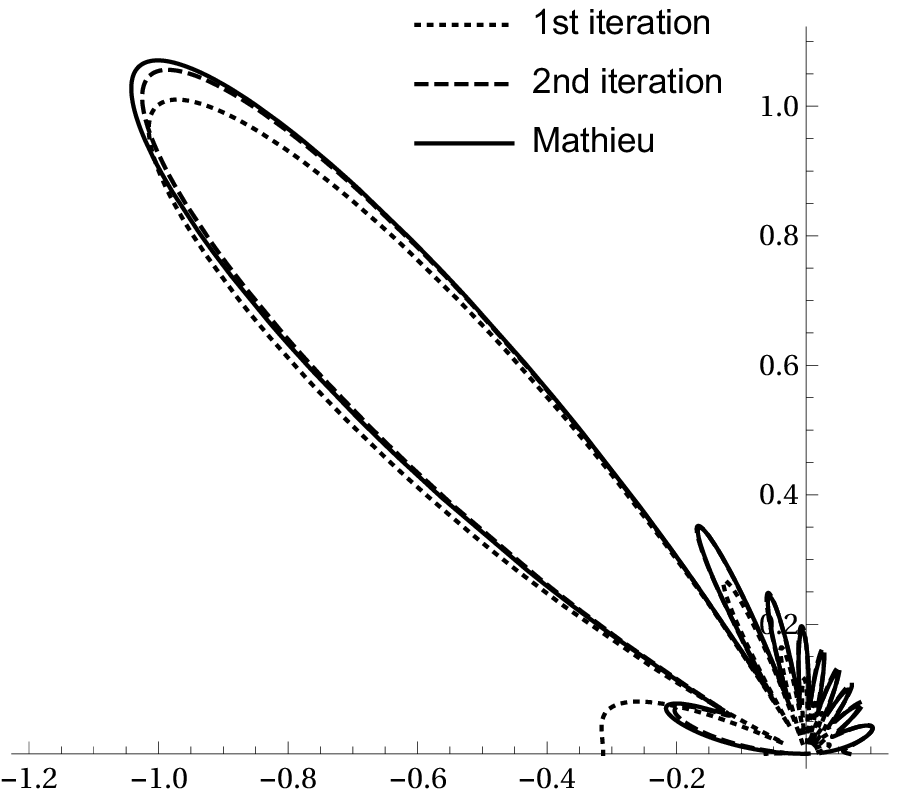}
\caption{$k_{0}=16$.\vspace{5pt}}
\end{subfigure}
\hfill
\begin{subfigure}[t]{0.5\linewidth}
 \centering
 \includegraphics[scale=0.65]{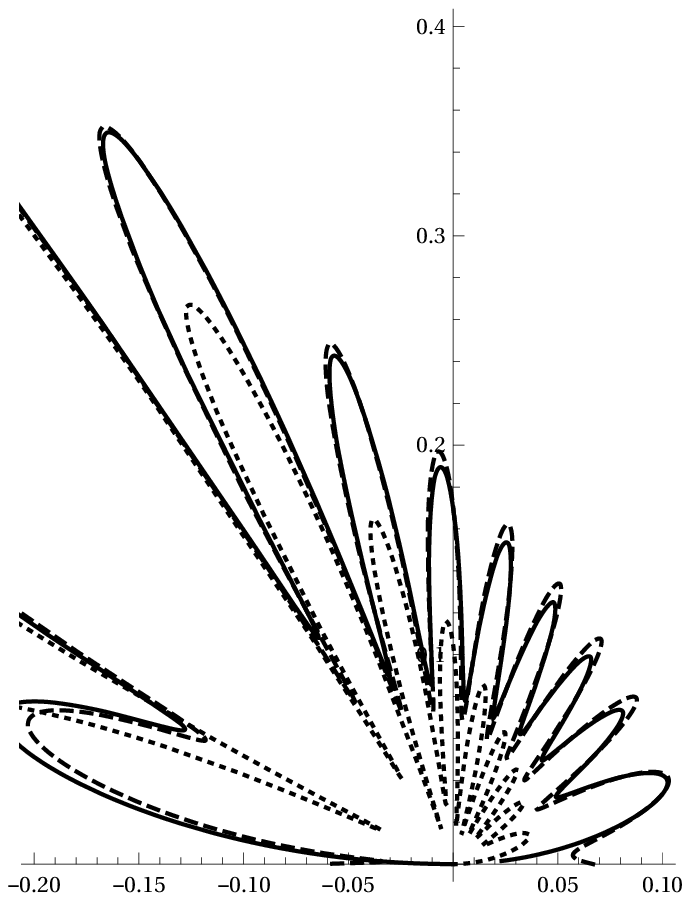}
\caption{$k_{0}=16$ view of highly oscillatory region.}
\end{subfigure}
\caption{Far-field scattered acoustic directivity generated by a plane sound wave incident at $45^{\circ}$ from the far field of varying frequencies $k_{0}$ scattering off of a finite impermeable plate of length $2$. Solutions obtained from the 1st and 2nd iterative steps are plotted against the Mathieu solution.}
\label{fig:finite}
\end{figure}

With this in mind we consider the scattering of plane waves of frequencies $k_{0}=1,6,
10, 16$. In Figure \ref{fig:finite} we show the truncated Mathieu
solutions against the 1st and 2nd iterative solutions for the
far-field acoustic directivity for a sound wave with incident angle
$\theta_{i}=\upi/4$. 
For $k_{0}=1$ we do not expect convergence of the result, as the initially neglected term is not small. Indeed in Figure \ref{fig:finite1} we see the first and second iterative solutions vary greatly, however they give us some insight into how the iterative procedure works: the initial solution neglects one scattering edge (at $x=0$), thus we see a semi-infinite style directivity pattern which is zero at $\theta=0$ but non-zero at $\theta=\upi$. The second iteration introduces a correction for the $x=0$ edge, thus forcing the pressure to be zero at $\theta=\upi$, but as the iterative procedure is not complete we no longer have zero pressure at $\theta=0$. Even after 5 iterations we found that the solution had not converged for $k_{0}=1$, thus we believe in this case the iterative procedure is simply not appropriate.

For frequencies $k_{0}=6, 10, 16$ there is good agreement between the iterative solution and the Mathieu solution after just 2 iterations, with better agreement for the higher frequencies (as expected). The key differences between the 1st and 2nd iterative solutions follow similar reasoning to the $k_{0}=1$, with the 1st iterative solutions having significant errors for $\theta=\upi$ which are quickly corrected by the time of the 2nd iteration. We note however that the first iteration does not resemble high-frequency single-edge scattering (i.e. a semi-infinite plate result) as it includes oscillations which arise from leading- and trailing-edge interaction. These are present in the first iterative solution because the forcing, $f_{3,4}$, used in the initial steps come from finite-plate forcing, not semi-infinite plate forcing, thus includes some interacting leading- and trailing-edge terms.

\subsection{Semi-infinite impermeable plate with finite porous extension}\label{sec:4b}

\subsubsection{Comparison to \cite{AYTON_PE}}
Here we compare results for our semi-infinite partially porous plate to the sound scattering results for a finite partially porous plate given by \citet{AYTON_PE}. \citet{AYTON_PE} uses a pole removal technique to obtain the far-field acoustic directivity due to a sound wave scattering off a flat plate that consists of a finite impermeable rigid section and a finite poroelastic section (which for the purposes of this section we consider to be only porous). For our new iterative method we require a multiplicative factorisation of $(\gamma P-1)$, equivalently,  $\mu + \gamma$, where the constant porosity parameter, $\mu$, is small. We can therefore use the asymptotic factorisation from \citet{CriLepp};
\begin{equation*}
 (\mu+\gamma(\alpha))_{\pm}=\gamma_{\pm}(\alpha)\left(1+\frac{\mu}{\upi\gamma(\alpha)}\cos^{-1}\left(\pm\frac{\alpha}{k_{0}}\right)+o(\mu)\right),
\end{equation*}
and neglect the $o(\mu)$ contribution. This factorisation is obtained by supposing $(\mu+\gamma)_{\pm}=\gamma_{\pm}+\mu G_{\pm} +o(\mu^{2})$, and using the known additive factorisation  $\gamma=\gamma^{+}+\gamma^{-}$ from \citet[p. 21]{Noble} to solve for $G_{\pm}$.

For investigation of $\mu=O(1)$ a numerical factorisation of $\mu+\gamma$ could be performed similar to that used in \citet{JP}. Note the pole removal method does not rely on an asymptotic factorisation thus can investigate plates with higher porosity distributions.

\begin{figure}
 \centering
 \includegraphics[scale=0.8]{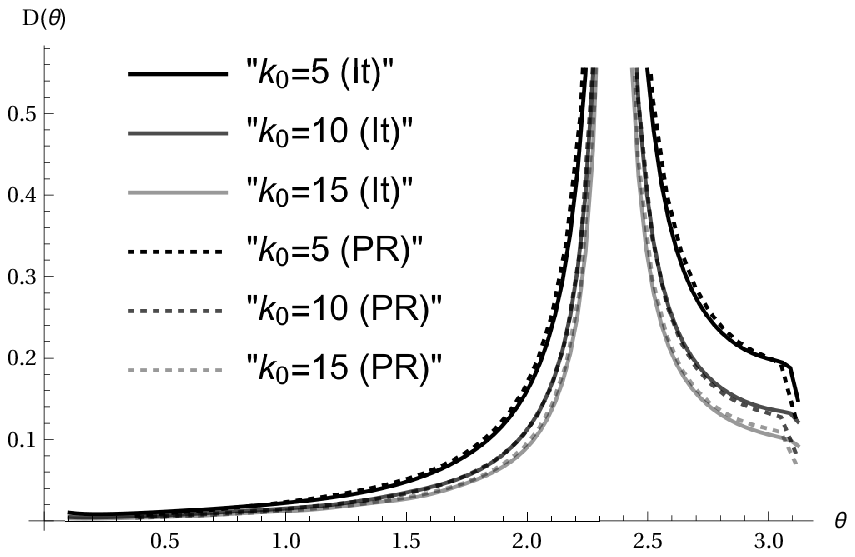}
\caption{Comparison of far-field directivity calculated using the iterative method (It) or pole removal method (PR) for an incident sound wave with frequency $k_{0}$ impinging on a semi-infinite flat plate with porous section of length $L=1$, $\mu=0.1$.}
\label{fig:semiinfsound}
\end{figure}

\begin{figure}
\begin{subfigure}[b]{0.45\linewidth}
 \centering
 \includegraphics[scale=0.7]{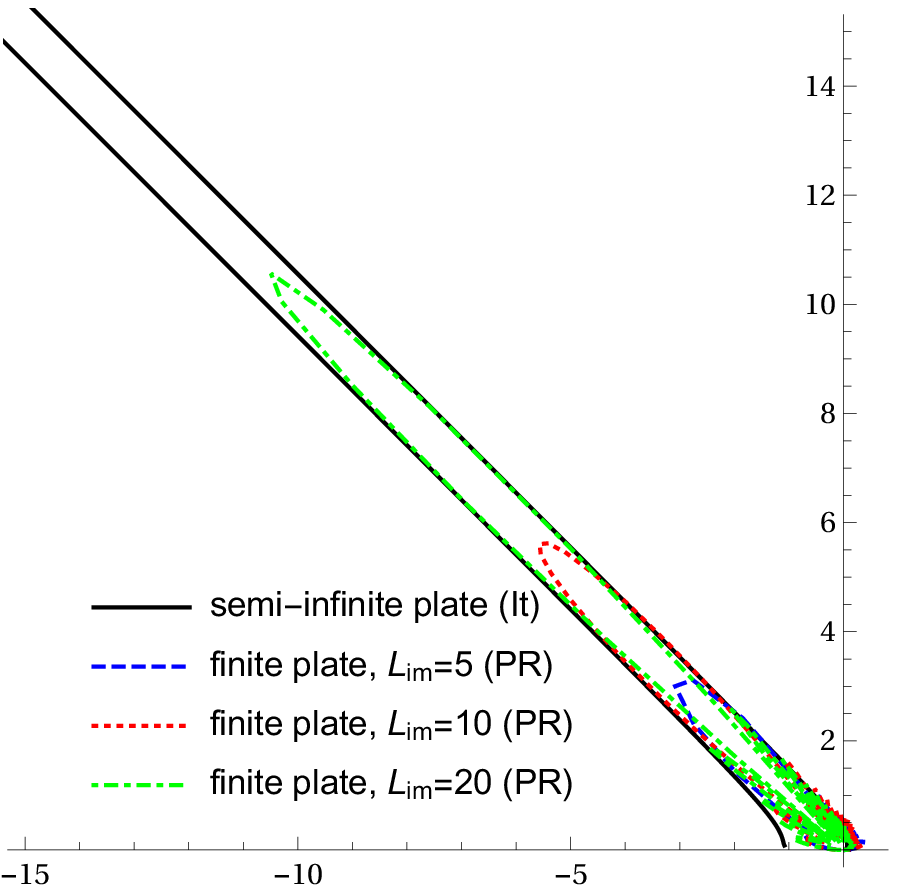}
\caption{Full range}
\end{subfigure}
\hfill
\begin{subfigure}[b]{0.45\linewidth}
 \centering
 \includegraphics[scale=0.7]{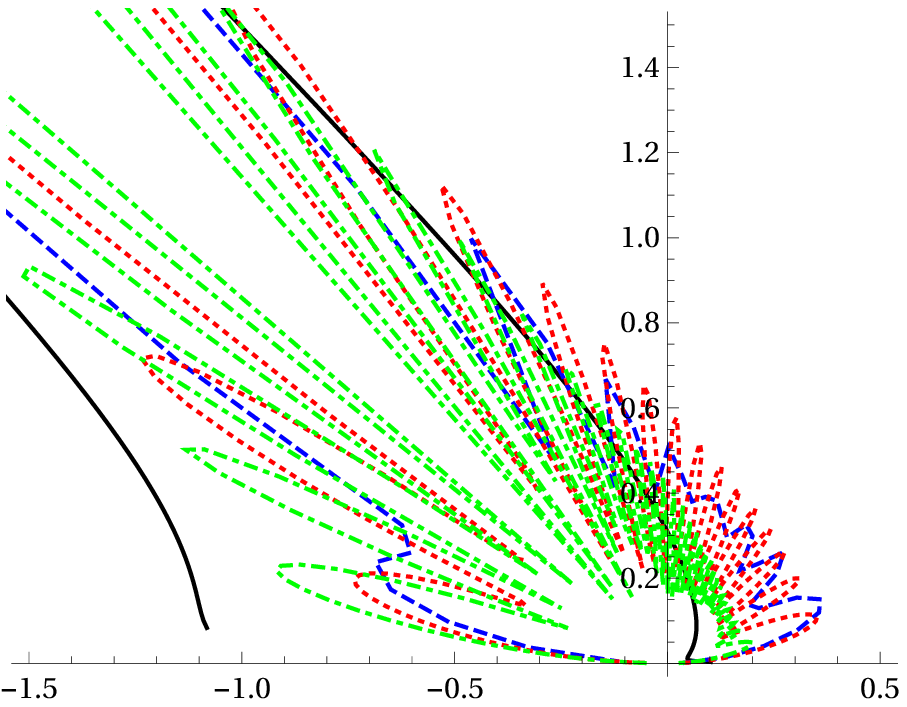}
\caption{Highly oscillatory region}
\end{subfigure}
\caption{Comparison of far-field directivity calculated using the iterative method (It) for a semi-infinite plate to the directivity from a finite plate calculated using the pole removal method (PR). Both have porous lengths of $L=1$, sound wave frequency $k_{0}=5$ incident from angle $\upi/4$, and porosity $\mu=0.1$. The finite impermeable plate length, $L_{im}$, is allowed to vary. (Colour online).}
\label{fig:finitecompare}
\end{figure}

First we consider a direct comparison of the two methods by applying the pole removal technique to the semi-infinite partially porous plate. In Figure \ref{fig:semiinfsound} we see good agreement between the two methods for mid and high frequencies. The results are heavily dominated by the Fresnel zone at the angle of reflection of the incident sound wave, $\theta=3\upi/4$ and uniformly valid expansions for the far-field directivity are required to deal with the pole singularity at this angle.
Note that the pole removal method of \citet{AYTON_PE} was seen only to be accurate at high frequencies and does not have a formal error bound to an exact solution, whilst our new iterative method is proved to converge to the exact solution \citep{My_iter} thus is able to present results to within a specified error bound. Further, the pole removal method can only recover the far-field acoustic field and not the mid or near field. A benefit of the new method in this paper is that we can not only consider a wider range of frequencies, but we can also obtain the scattered field at any point in the domain, thus can use near-field results from a far-field incident sound wave to obtain an acoustic amplification factor for far-field scattering from a near field source by reciprocity \citep{JP}. We shall comment on this amplification factor shortly. The main advantage of the pole removal method over the iterative approach is that it is easy to adapt for use on any size matrix Wiener-Hopf problem (\citet{AYTON_PE} considers a 3 x 3 matrix problem), whilst the iterative method would be more complicated to adapt to higher dimensions, and likely require stricter conditions to be placed on the parameters used to ensure convergence.

By including the finite leading edge we see it have a big effect on the overall directivity in Figure \ref{fig:finitecompare} for the scattering of a sound wave. In particular the Fresnel regions at the angle of direct reflection of the incident wave (thereby indicated by the largest lobes) are significantly altered by the extra leading-edge scattered field. Uniformly valid expansions for the far-field directivity are now \emph{not} required, as the combination of the Fresnel regions from the leading and trailing edges regularise one another, resulting in the pole at $\theta=3\upi/4$ becoming a removable singularity. As such, the leading-edge field significantly reduces the overall impact of the trailing-edge Fresnel region on the far-field noise. As the length of the impermeable section is increased, the amplitude of the Fresnel region increases, and the modulation of the directivity increases. We also observe that the finite leading edge reduces the scattered field directly upstream for any plate length, illustrating that the leading-edge field is formally the same magnitude as the trailing-edge field close to the downstream direction (as discussed in Section \ref{sec:4a}).

For sound scattering we therefore note that there is a great influence of a finite leading edge on the overall predictions of the far-field noise as both leading and trailing-edge scattered fields particularly in the Fresnel regions since both the leading and trailing edge directly scatter the sound wave at angle $\theta=3\upi/4$. However, we are not concerned by this influence for predicting trailing-edge noise as our convective gust model supposes the incident field does not interact with the leading edge directly. And, of course, for a convective gust there are no Fresnel regions thus no regularisation of the far-field directivity is required. 

\subsubsection{Trailing-edge noise}

Now that we have validated our new approach, we consider the original problem of a convective boundary layer gust interacting with the semi-infinite partially porous plate. In all results in this section we shall take $\theta_{i}=\upi/4$, and $L=1$.

\begin{figure}
 \centering
\begin{subfigure}[b]{0.45\linewidth}
 \centering
 \includegraphics[scale=0.6]{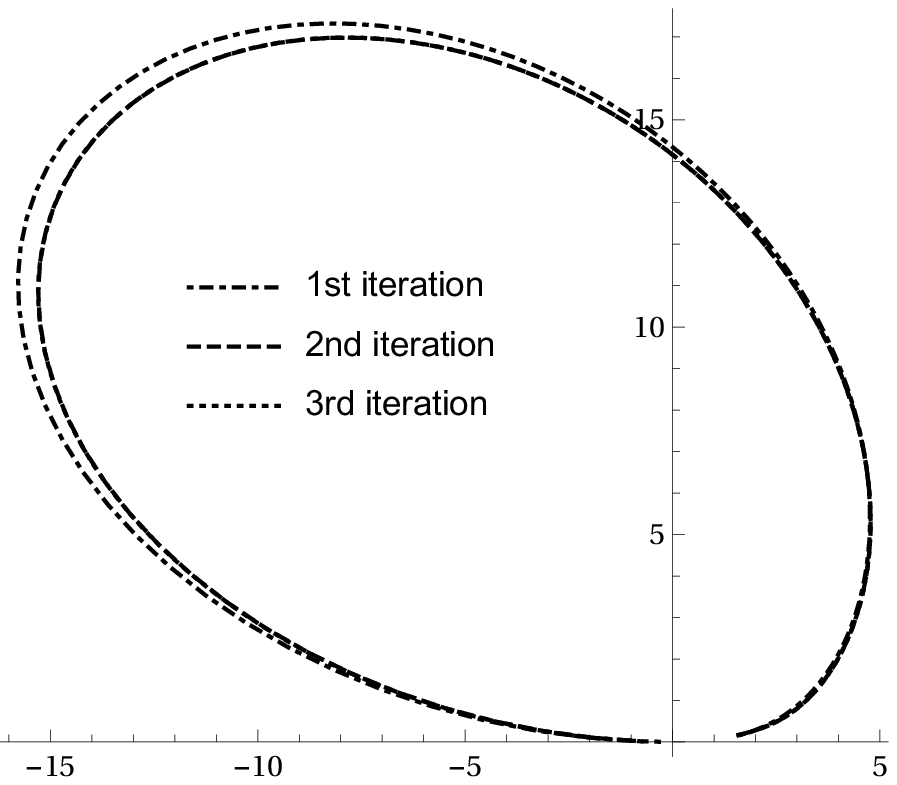}
\caption{$k_{0}=0.1$, $\mu=0.1$.\vspace{5pt}}
\end{subfigure}
\begin{subfigure}[b]{0.45\linewidth}
 \centering
 \includegraphics[scale=0.6]{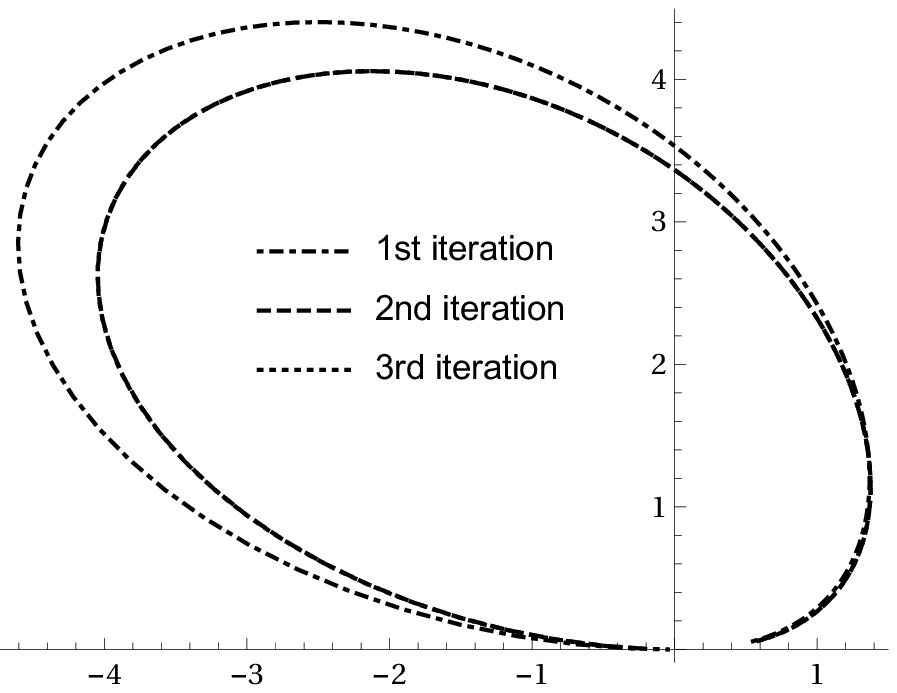}
\caption{$k_{0}=0.5$, $\mu=0.3$.\vspace{5pt}}
\end{subfigure}
\\
\begin{subfigure}[b]{0.45\linewidth}
 \centering
 \includegraphics[scale=0.6]{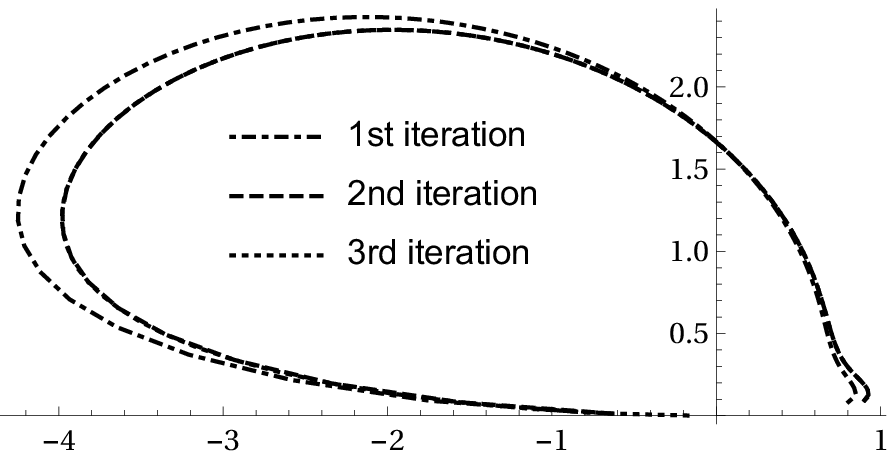}
\caption{$k_{0}=1$, $\mu=0.1$.\vspace{5pt}}
\end{subfigure}
\begin{subfigure}[b]{0.45\linewidth}
 \centering
 \includegraphics[scale=0.6]{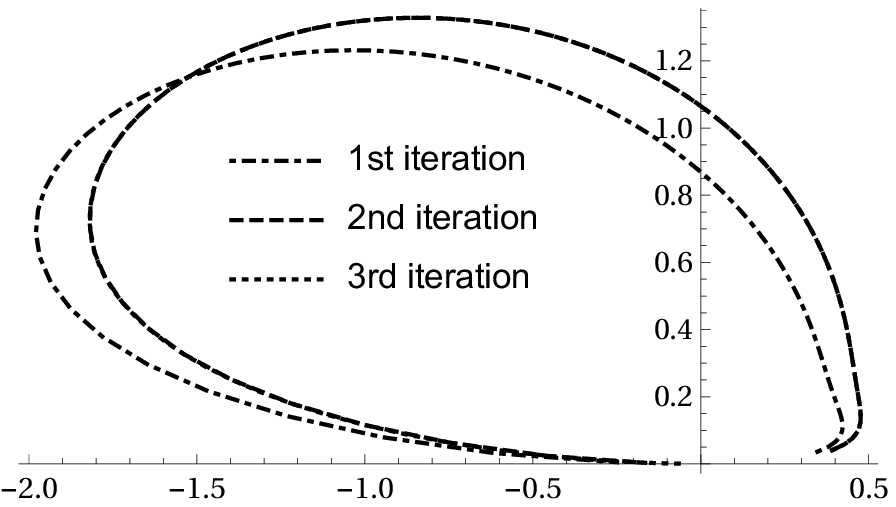}
\caption{$k_{0}=2$, $\mu=0.3$.\vspace{5pt}}
\end{subfigure}
\caption{Far-field acoustic directivity generated by a convective gust in mean flow $M=0.6$ of varying frequencies $k_{0}$ interacting with a semi-infinite impermeable plate with an attached finite porous plate of unit length and porosity parameter $\mu$. First, second, and third iterative solutions are presented.}
\label{fig:convergenceP}
\end{figure}

\begin{figure}
 \centering
\begin{subfigure}[b]{0.45\linewidth}
 \centering
 \includegraphics[scale=0.7]{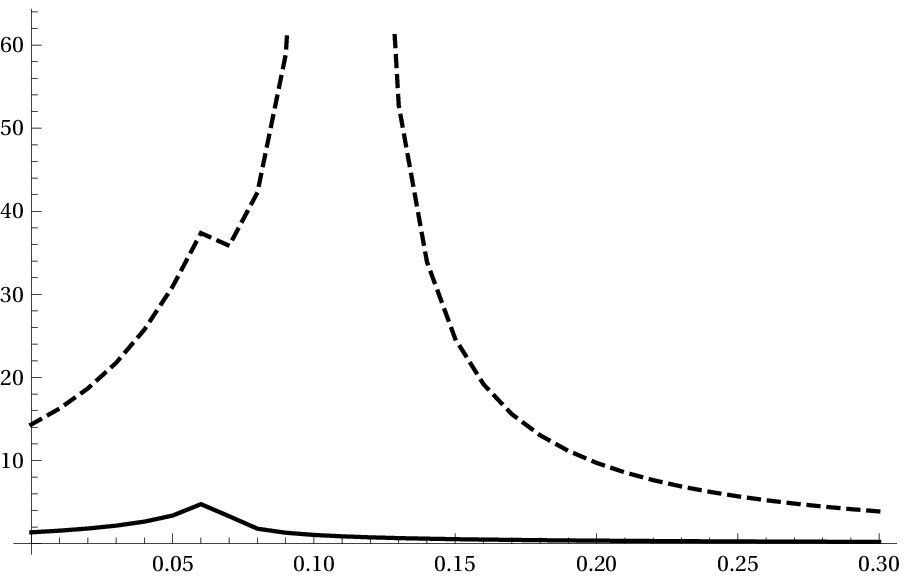}
\caption{$k_{0}=0.1, \mu=0.1$.\vspace{5pt}}
\end{subfigure}
\hspace{5pt}
\begin{subfigure}[b]{0.45\linewidth}
 \centering
 \includegraphics[scale=0.7]{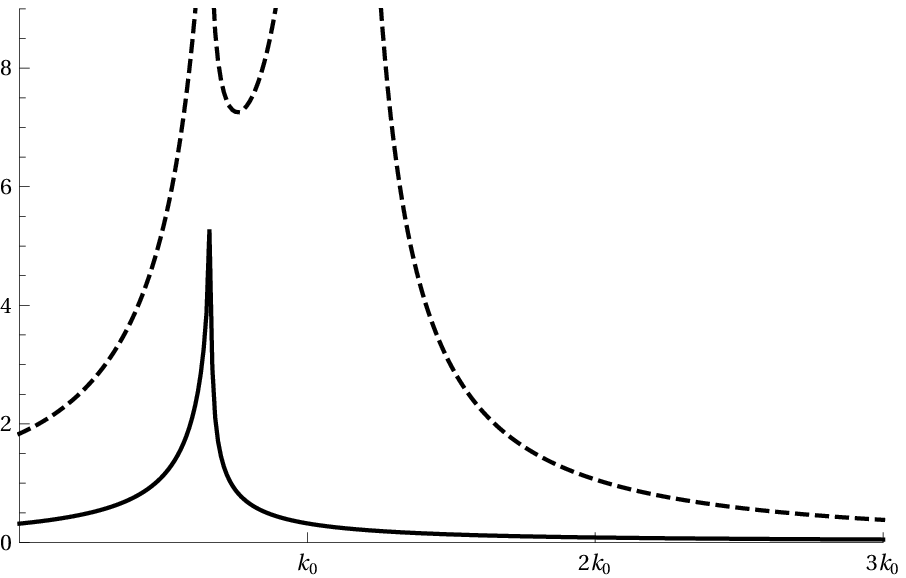}
\caption{$k_{0}=1, \mu=0.3$.\vspace{5pt}}
\end{subfigure}
\caption{Absolute value of the two terms on the RHS of \eqref{eq:eq3} as calculated after one complete iterative loop. First term is the dashed line, second term (which is initially neglected) is the solid line.}
\label{fig:neglected}
\end{figure}

We begin by illustrating the convergence of the iterative procedure in Figure \ref{fig:convergenceP} which shows the far-field acoustic directivity for different $k_{0}$ and $\mu$ values. After 2 iterations the solution is indistinguishable from the 3rd iteration in each case with differences of less that $10^{-5}$ between the two solutions.
The convergence of the iterative method is fast in this case for low
frequency (compared to the finite plate case where the condition of
high frequency ensured quick convergence), since we restrict attention
to low porosity, and therefore by considering the scalings of terms in
the initial equation, \eqref{eq:eq3}, in terms of porosity, $\mu\ll1$,
or equivalently $|P|\gg1$, we see the initially neglected term,
$\Phi_{-}^{(0)}\e^{-\i\alpha L}$, is smaller than those retained. We
illustrate the relative size of the neglected terms in Figures
\ref{fig:neglected}  to verify it is indeed formally smaller than the
initially retained terms, thus we expect (and see) quick convergence
of the iterative method. For lower values of $\mu$ and lower
frequencies $k_{0}$ the relative size of the neglected term is much
smaller than at higher values of $\mu$ and $k_{0}$, however in these
cases the neglected term is still formally smaller and therefore the
iterative procedure converges. 

From hereon in our results are given as the 2nd iterative solution. 

\begin{figure}
 \centering
\begin{subfigure}[b]{0.45\linewidth}
 \centering
 \includegraphics[scale=0.7]{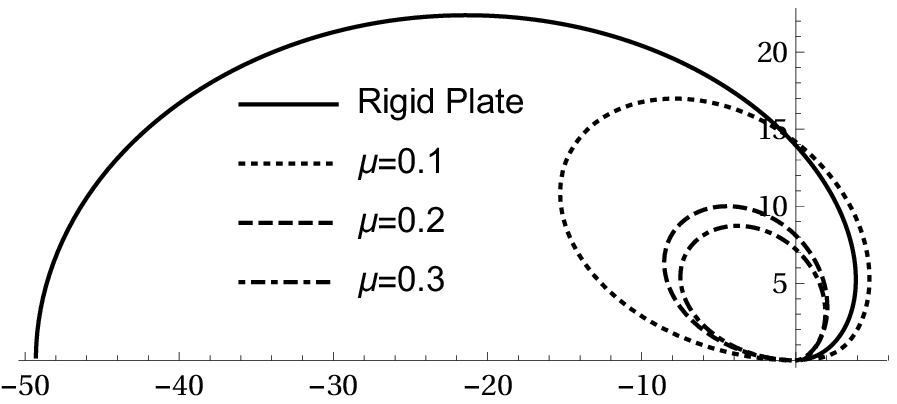}
\caption{$k_{0}=0.1$.\vspace{5pt}}
\end{subfigure}
\hfill
\begin{subfigure}[b]{0.45\linewidth}
 \centering
 \includegraphics[scale=0.7]{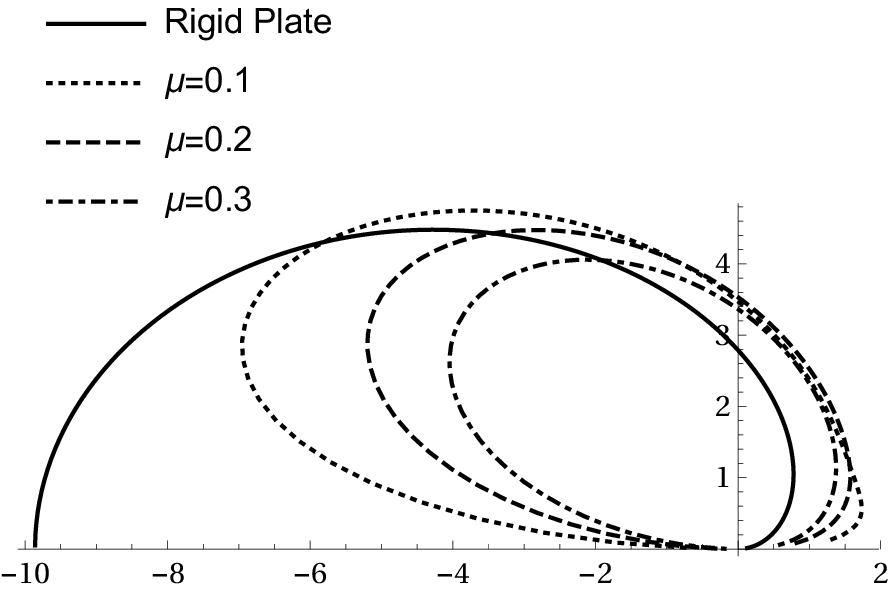}
\caption{$k_{0}=0.5$.\vspace{5pt}}
\end{subfigure}
\begin{subfigure}[b]{0.45\linewidth}
 \centering
 \includegraphics[scale=0.7]{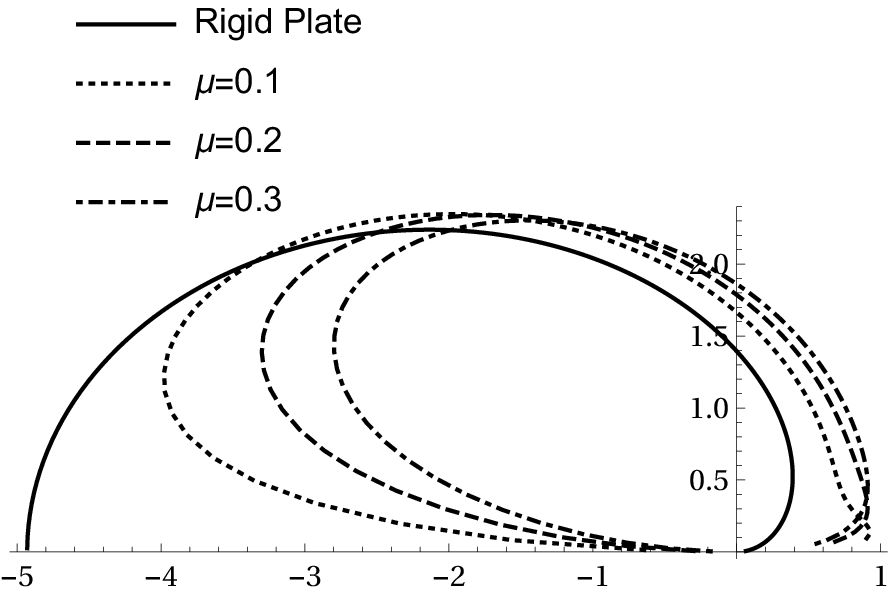}
\caption{$k_{0}=1$.\vspace{5pt}}
\end{subfigure}
\hfill
\begin{subfigure}[b]{0.45\linewidth}
 \centering
 \includegraphics[scale=0.7]{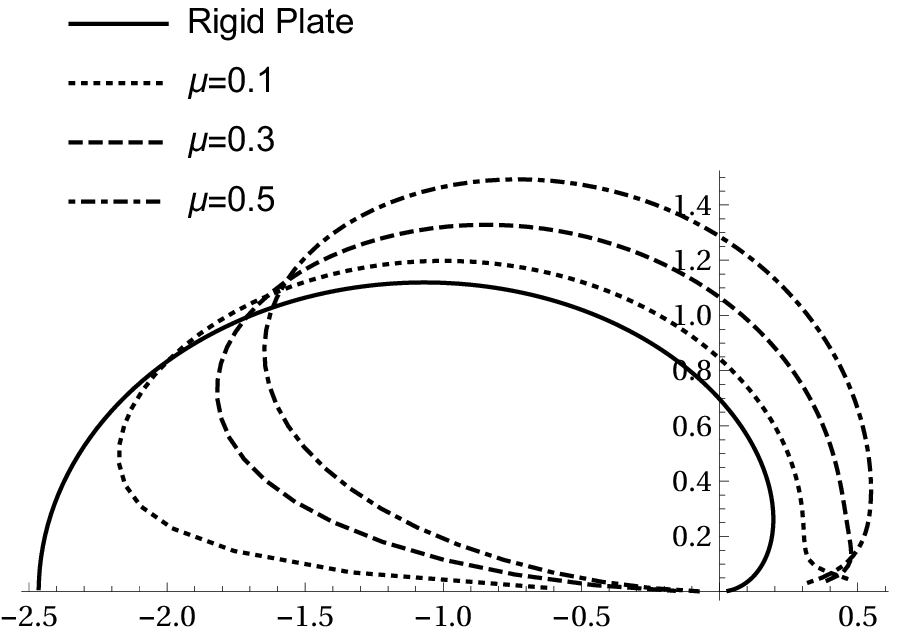}
\caption{$k_{0}=2$.\vspace{5pt}}
\end{subfigure}
\caption{Far-field acoustic directivity generated by gusts of varying frequencies $k_{0}$ interacting with a semi-infinite impermeable plate with an attached finite porous plate of unit length and porosity parameter $\mu$, in mean flow $M=0.6$.}
\label{fig:porous}
\end{figure}

Figure \ref{fig:porous} shows the effects of altering the porosity parameter $\mu$ on the far-field acoustic  directivity for gusts of different frequencies, $k_{0}$. We see that the reduction of far-field acoustic pressure achievable by an increasingly porous plate is lessened as incident frequency increases as expected from the results of \citet{JP}. However at high frequency the directivity pattern, although not significantly reducing in magnitude, significantly changes in overall shape.

For low frequencies, the acoustic directivity behaves as a single source from just the trailing edge, $x=L$, (driven by a pressure jump across the trailing edge) therefore increasing porosity behaves in an identical manner to that seen for the scattering of a near-field trailing-edge source by a finite porous plate investigated in \citet[Figure 7a]{CavaWolf}. 
At higher frequencies, the source is no longer compact therefore there is notable interaction from scattering by the trailing edge, $x=L$, and the impermeable-permeable junction, $x=0$. This secondary acoustic field, emanating from the permeable-impermeable junction, is expected to be weaker than the secondary source in \citet{CavaWolf}, the (finite) leading-edge field. We see from Figure \ref{fig:porous} that at high frequency $k_{0}=2$, increasing porosity rotates the directivity away from the upstream direction without drastically altering the overall magnitude compared to the rigid plate. This effect is different to the effects of the finite leading edge seen in \citet[Figure 7b]{CavaWolf} (reduction of overall magnitude and oscillations).

We can consider the effect of the secondary acoustic field emanating from permeable-impermeable junction by considering the correction term between the first iterative solution and the converged iterative solution (which for the cases we consider here is the second iterative solution), since the first solution accounts for the edge at $x=L$ and successive solutions correct for the edge at $x=0$. 
Figure \ref{fig:convergenceP} shows the far-field directivities calculated by the first and second iterative solutions; for low frequencies this correction rescales the overall pressure almost uniformly over all $\theta$ directions. This indicates the source at $x=0$ behaves in an identical but weaker manner to the source at $x=L$, i.e. is dipole-like. 
For $k_{0}=2$, $\mu=0.3$ we see a significant change in directivity between the first and second iterative solutions, therefore the permeable-impermeable junction is no longer behaving in an identical manner to the trailing edge. The trailing edge (first solution) directs noise strongly at some angle $\upi/2<\vartheta<\upi$ as expected for edge interaction in subsonic flow with high Mach number, $M=0.6$. The permeable-impermeable junction (correction to the first solution) instead seems to direct noise predominantly at some angle $\vartheta'<\vartheta$.

\begin{figure}
 \centering
\begin{subfigure}[b]{0.45\linewidth}
 \centering
 \includegraphics[scale=0.7]{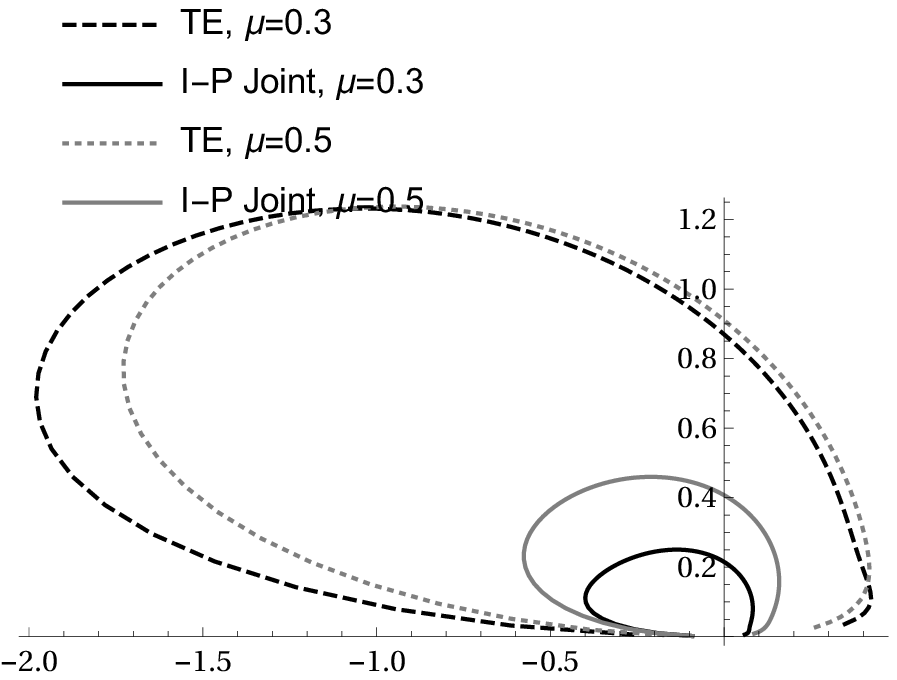}
\caption{$k_{0}=2$.\vspace{5pt}}
\end{subfigure}
\hspace{5pt}
\begin{subfigure}[b]{0.45\linewidth}
 \centering
 \includegraphics[scale=0.7]{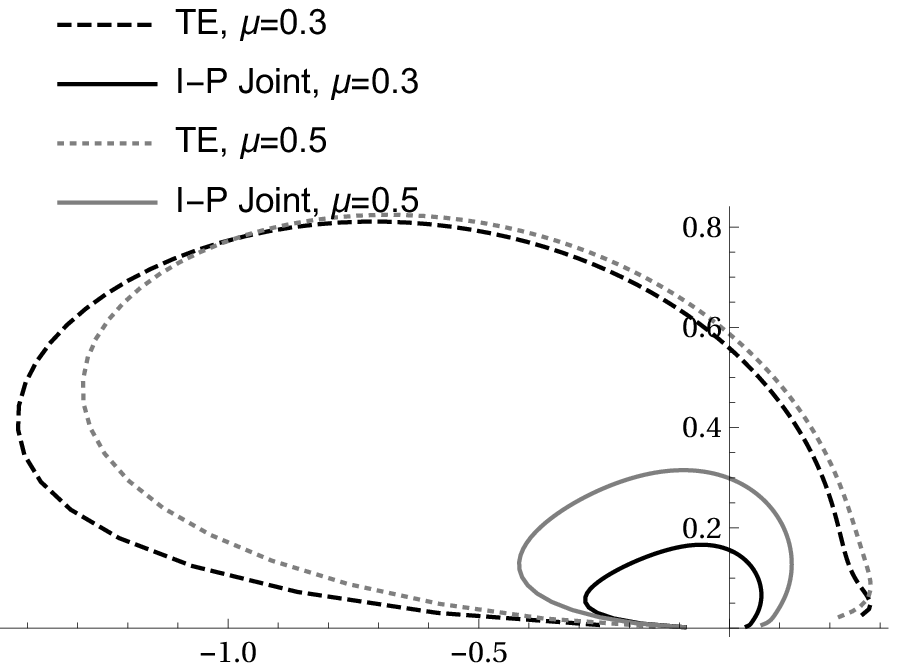}
\caption{$k_{0}=3$.\vspace{5pt}}
\end{subfigure}
\caption{Absolute value of the difference between the first and second iterative solutions (I-P Junction), compared to the first iterative solution (TE) for $M=0.6$.}
\label{fig:diff}
\end{figure}

In Figure \ref{fig:diff} we investigate the far-field directivity
generated by the difference between the first and second iterative
solutions for high-frequency interactions with $M=0.6$, and indeed see
that for higher frequencies the impermeable-permeable correction term
has a significantly different directivity pattern that is directed
away from the main region of pressure for the trailing edge term. We
also see that whilst increasing porosity reduces the magnitude of the
trailing-edge term, it increases the magnitude of the correction term,
therefore when the two terms are combined there is little difference
in the overall magnitude of the acoustic pressure in Figure
\ref{fig:porous} (d). The directivity pattern generated by the correction term is significantly different to the $\sin\theta/2$ directivity expected from a finite leading edge.


\begin{figure}
 \centering
 \includegraphics[scale=0.8]{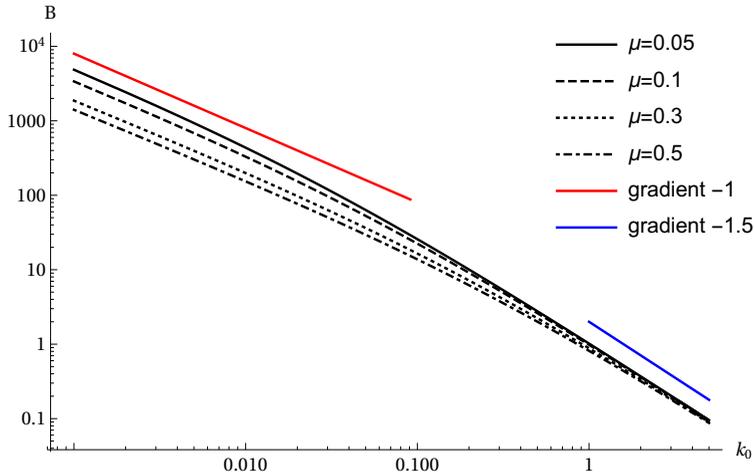}
\caption{Amplification factor, $B$, as a function of frequency $k_{0}$ for a range of partially porous plates. (Colour online).}
\label{fig:amplification}
\end{figure}

Finally we can investigate the acoustic amplification factor, $B$ due to a near-field acoustic source; by considering the near-field scattering of a far-field incident sound wave we can invoke reciprocity to obtain the far-field acoustic amplification due to a near-field quadrapole source. The amplification factor is defined in \citet{JP} as $B=|\tilde{B}|k_{0}^{-2}$, where
\begin{equation}
 \pderiv{\Phi_{+}}{y}(\alpha,0)\sim \tilde{B} \alpha^{-1/2}\qquad \text{as}\,\,\,|\alpha|\to\infty.
\end{equation}
It is seen in \citet{JP} that for a low porosity edge with $\alpha_{H}=0.0014$ (equivalent to $\mu=0.034$) the amplification factor as a function of frequency has a gradient of $-1$ for low frequency, and a gradient of $-1.5$ for high frequency (which is also the gradient found for a rigid edge).
We recover these limits for low porosities in Figure \ref{fig:amplification} as expected since the near-field quadrapole primarily only interacts with the porous section of the plate, thus this can be well approximated by the semi-infinite porous plate. 

These results further validate our new approach, however considering a near-field trailing-edge source does not suitably investigate the effects of a impermeable-permeable junction lying within the turbulent boundary layer -  this paper has focussed on a convective gust over both the impermeable and permeable sections of the plate to enable turbulence to interact with both the junction and the trailing edge.

In this section we have determined that at low frequencies the impermeable-permeable junction seems not to greatly affect the noise generated in the far field by near field turbulence, and thus the predictions by \citet{JP} using a semi-infinite permeable plate are accurate even if the plate were only partially porous. However, for higher frequency turbulence, the impermeable-permeable junctions has a greater effect on the far-field directivity pattern, thus we expect for high frequencies the predictions of \citet{JP} may not be sufficiently accurate to present the full picture of a partially porous aerofoil. Thus we investigate the effects of the impermeable-permeable junction on specifically high-frequency interactions in the following section.

\subsection{Effects of porosity on high-frequency turbulence}
It is known that trailing-edge noise dominates overall aerofoil noise for high-frequency interactions, whilst at mid- to low-frequency interactions leading-edge noise dominates \citep{Chai}. A proposed parameter indicating where trailing-edge noise becomes dominant is $\omega^{*}t/U^{*}>1$, where $\omega^{*}$ is dimensionalised frequency, $t$ is the ratio of aerofoil thickness to chord length, and $U^{*}$ is dimensionalised mean flow velocity. Thus if we wish to investigate the effects of porosity on total aerofoil noise generation, we wish to consider a high-frequency regime where trailing-edge noise plays a key part. Despite previous studies indicating that porosity only has an effect at low frequencies, \citep{JP}, we have seen in the previous section that a partially porous semi-infinite plate behaves differently to a fully rigid (or indeed fully porous) semi-infinite plate, due to the interaction between the impermeable-permeable junction and the trailing edge tip itself. This interaction is influential at high frequencies because the wavelength of the gust is much shorter than the distance between the two points, and is clearly not modelled in \citet{JP}'s work for a fully porous plate.

\begin{figure}
 \centering
 \includegraphics[scale=0.8]{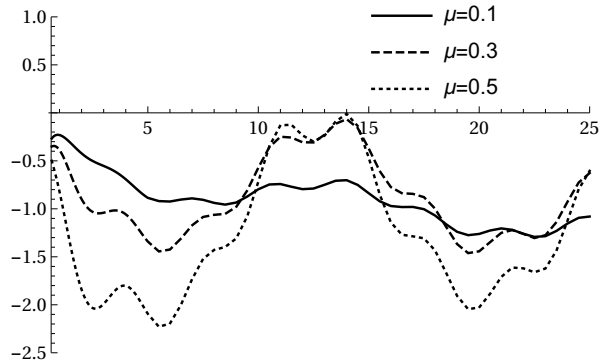}
\caption{$\Delta P(k_{0})$ for semi-infinite partially porous plates with varying porosity parameter, $\mu$.}
\label{fig:rigidporouscompare}
\end{figure}

Figure \ref{fig:rigidporouscompare} shows the difference in dB of the sound power predicted from a fully porous plate to that predicted from our partially porous plate, namely
\begin{equation}
 \Delta P (k_{0})= 10\log_{10}\left[\frac{\int_{0}^{\upi}|D_{\text{fully}}(\theta)|^{2}d\theta}{\int_{0}^{\upi}|D_{\text{partial}}(\theta)|^{2}d\theta}\right],
\end{equation}
where $D_{\text{partial}}(\theta)$ is the directivity predicted for our partially porous plate, \eqref{eq:directivity}, and $D_{\text{fully}}(\theta)$ is the corresponding directivity predicted if the plate were fully porous. 

\begin{figure}
 \centering
 \includegraphics[scale=0.8]{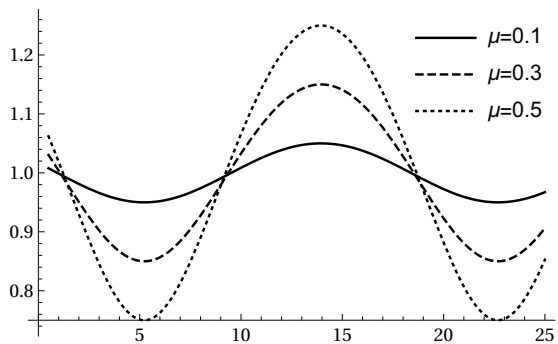}
\caption{Interference function, $1+\mu\, \e^{\i M^{2} L (k_{0}+3)}$, for various $\mu$ with fixed $L=1$.}
\label{fig:interference}
\end{figure}

We see for all mid to high frequencies and all porosity parameters the relative difference is always negative. We therefore assess that the effect of the permeable-impermeable joint is to slightly increase the total noise generated at the trailing edge in comparison to either a fully rigid or fully porous plate (since at high frequencies these two predictions are identical). This increase is minor (approximately 1-2dB) in comparison to the decreases in noise seen at low frequencies. The increase is greatest for higher porosity plates since there is a bigger difference in the two scattering points thus a greater interaction effect of the two fields. The difference in sound oscillates with frequency $k_{0}$ since an interference between the trailing edge and the impermeable-permeable junction can be constructive or destructive. Our prediction of an increase in noise for a partially porous plate is in agreement with the experimental findings of \citet{GeyerPorous}, although they speculated that their observed high-frequency noise increase for partially porous aerofoil (compared to fully impermeable aerofoils) was due to increased surface roughness noise. As our model does not allow for roughness noise, we propose a combination of increased roughness noise, and interference effects between the impermeable-permeable junction and trailing edge can lead to an increase of noise at high frequencies when an aerofoil is partially porous.

Using the results of Figure \ref{fig:rigidporouscompare} we can estimate the interaction effect of the impermeable-permeable junction to be of the form $1+\mu \,\e^{\i M^{2} L k_{0} }$, where the $1$ denotes the scattering from the porous trailing-edge tip, and the $\mu \,\e^{\i M^{2}  L k_{0}}$ is the interference caused by the impermeable-permeable junction, which sensibly tends to zero as the plate becomes fully rigid. In Figure \ref{fig:interference} we plot this interaction function (shifted to align with Figure \ref{fig:rigidporouscompare}) across the same range of frequencies as used Figure \ref{fig:rigidporouscompare}, and see good relative agreement between results. This implies that the overall radiation is generated by two fields; a trailing-edge field and an impermeable-permeable junction field which interfere when summed in the far field.

\begin{figure}
\begin{subfigure}[b]{0.45\linewidth}
 \centering
 \includegraphics[scale=0.7]{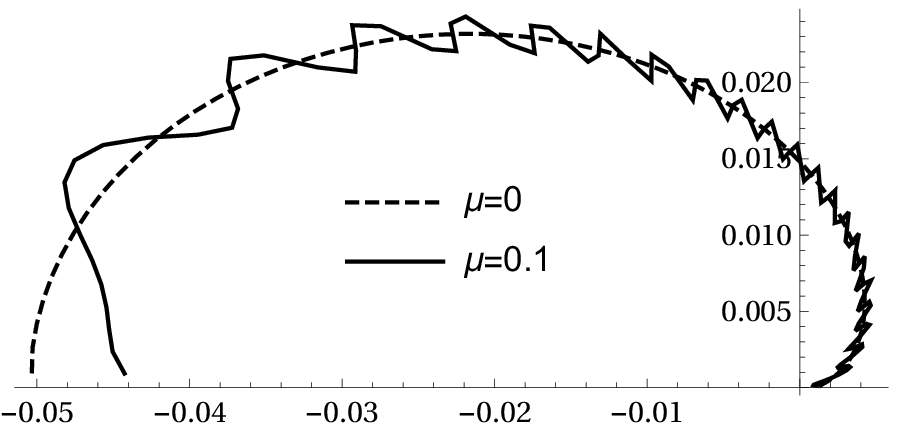}
\caption{$k_{0}=100$}
\end{subfigure}
\hfill
\begin{subfigure}[b]{0.45\linewidth}
 \centering
 \includegraphics[scale=0.7]{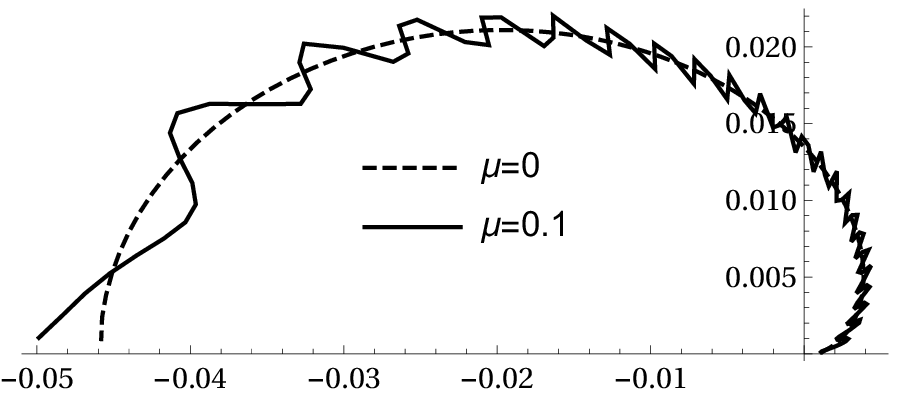}
\caption{$k_{0}=110$}
\end{subfigure}
\caption{Far-field directivity for partially porous plate with $\mu=0.1$, $L=1$, $M=0.6$ compared to a fully rigid plate, $\mu=0$.}
\label{fig:k0100}
\end{figure}

One would perhaps expect as $k_{0}\to\infty$ the interaction effect to significantly decay as the interference should scale inversely with $k_{0}$ and therefore the partially porous plate to behave identically to the fully porous plate. We do not see this in our integrated results because the two acoustic fields directly upstream ($\theta=\upi$) are of the same magnitude thus any constructive or destructive interference here will always have a notable effect on the total sound. We illustrate this feature, in the extreme cases of $k_{0}=100, 110$, in Figure \ref{fig:k0100}; we clearly see for the majority of the far field the acoustic directivities are in very good agreement, however as we approach the downstream direction, the interference leads to a large different between the directivities.

\section{Conclusions}\label{sec:5}

This paper investigates the interaction of a convective gust with a partially porous semi-infinite plate in uniform steady flow. We find, in agreement with previous work, that porous trailing-edge adaptations can significantly reduce low-frequency interaction noise. New insight into the effects of a the impermeable-permeable junction have been found as we see increasing porosity has the capability to rotate the far-field directivity away from the upstream region for mid-frequency interactions due to interference between acoustic fields from the permeable-impermeable junction, and the trailing-edge tip itself. Whilst a porous trailing-edge adaptation does not significantly reduce far-field mid-frequency trailing-edge noise it could prove useful for controlling the predominant direction of noise propagation. Further, at high-frequencies the interference of the impermeable-permeable junction has been seen to increase the total far-field noise slightly when compared to a fully rigid plate in agreement with experimental findings of \citet{GeyerPorous}.

Elasticity has not been included in this problem as the interaction between the mean flow and the elastic plate would cause significant complications to the analysis. An approach involving matched asymptotic expansions, similar to \citet{Abrahamsflow}, and the Wiener-Hopf technique could be used to tackle the elasticity problem in the future. We have also restricted the porosity of the extension to satisfy $\mu\ll1$ to allow both an asymptotic factorisation of the kernel function, $\mu+\gamma$, and to ensure rapid convergence of the iterative solution. This restriction only discounts a small range of values investigated by \citet{JP}. Currently the iterative procedure using an asymptotic factorisation of $\mu+\gamma$, implemented in Mathematica on a standard 4-core desktop computer, calculates the third iteration for the results presented in under 60 seconds.

This paper also presents the application of a new way of solving
Wiener--Hopf equations of~\eqref{eq:main}. We present a
constructive iterating method of approximate solution of the Wiener--Hopf matrix problem.  This is an innovative way
which allows us to consider the two scalar equations corresponding to the
transition points separately and then couple them in an iterative
manner. This paper provided some examples of how the
procedure could be numerically implemented in acoustics.    Both  problems considered cannot be solved using standard
Wiener--Hopf techniques. The solution of finite rigid plate problem is
obtained using a different method (Mathieu functions) and  is presented in order to
validate the proposed method. They were found to be in good
agreement. It is found that for the two acoustic problems which were
considered the convergence is very fast and only two or three iterations are
required. We have also included an illustration of when the method is
not expected to converge see Figure \ref{fig:convergenceP} (a). This
is due to the fact that the neglected terms where not small. We believe this method to be more versatile and more accurate than the pole removal method used in \citet{AYTON_PE}, as the iterative method can allow the solution to be found anywhere in the scattering domain (not just the far field), and is proven to formally converge to an exact solution, thus allows us to determine accurate error bounds.

We expect the proposed method to find applications in a variety of
fields. Many problems could be reduced to  Wiener--Hopf equations of
type~\eqref{eq:main}. The matrix is typical for problems which have two
changes in the boundary conditions. This method could also be applied
to a wider class of Wiener--Hopf systems than~\eqref{eq:main}, in
particular Wiener--Hopf matrix  need not be triangular.


\section{Acknowledgements}

The authors are very grateful to Prof. Nigel Peake for suggesting this
problem and useful discussions along the way.
A. Kisil acknowledges
support from Sultan Qaboos Research Fellowship at Corpus Christi
College at University of Cambridge. \\
L. Ayton acknowledges support from a Research Fellowship at Sidney Sussex College.

  \bibliographystyle{jfm}

\end{document}